\documentclass{aa}
\usepackage{graphicx,epsfig}

\begin{document}
   \title{Accretion disk wind in the AGN broad-line region: Spectroscopically
resolved line profile variations in Mrk\,110}

   \author{W. Kollatschny
          \inst{1,2}\fnmsep
          \thanks{Based on observations obtained with the Hobby-Eberly
               Telescope, which is a joint project of the University
               of Texas at Austin, the Pennsylvania State
           University, Stanford University, Ludwig-Maximilians-Universit\"at
            M\"unchen, and Georg-August-Universit\"at G\"ottingen.}
         \
%         , K. Bischoff\inst{1}, E.L. Robinson\inst{2}, and W.F. Welsh\inst{2,3} 
          }

   \offprints{W. Kollatschny}

   \institute{Universit\"{a}ts-Sternwarte G\"{o}ttingen,
              Geismarlandstra{\ss }e 11, D-37083 G\"{o}ttingen, Germany\\
              \email{wkollat@uni-sw.gwdg.de}
         \and
             Department of Astronomy and McDonald Observatory,
             University of Texas at Austin, Austin, TX 78712, USA}
%         \and
%             Department of Astronomy, San Diego State University,
%             San Diego, CA 92182, USA}

   \date{Received date, 2003; accepted date, 2003}

   \abstract{
Detailed line profile variability studies
of the narrow line Seyfert 1 galaxy Mrk\,110
are presented.
We obtained the spectra 
in a variability campaign carried out
with the 9.2m Hobby-Eberly Telescope at McDonald Observatory.
The integrated Balmer and Helium (He\,{\sc i}, {\sc ii})
emission lines 
are delayed by 3 to 33 light days to the optical 
continuum variations respectively. 
The outer wings of the line profiles
 respond
much faster to continuum variations 
than the central regions.
The comparison of the observed profile variations
with model calculations of different  velocity fields
indicates an accretion disk structure
of the broad line emitting region in Mrk\,110.
Comparing the velocity-delay maps of the different emission lines
among each other a clear radial stratification
in the BLR can be recognized.
Furthermore,
delays of the red line wings are slightly
shorter than those of the blue wings.
This indicates an accretion disk wind
in the BLR of Mrk\,110. We determine a central
black hole mass of 
M = $1.8\cdot10^{7} M_{\odot}$.
Because of the poorly known inclination angle of the accretion disk
this is a lower limit only.
   }
  
  \authorrunning{W. Kollatschny}
  \titlerunning{Accretion disk wind in Mrk\,110}
   \maketitle
%
%________________________________________________________________
%
\section{Introduction}
The central broad emission line region (BLR) in active galactic nuclei
is unresolved. But a study of the variable ionizing  
continuum source and the delayed response of the broad emission lines
provides indirect information about size and structure
of the line emitting region
and their internal kinematics.
In this paper we present a detailed study of 
continuum and emission line profile variations 
in the Seyfert 1 galaxy Mrk\,110. Our goal is 
to determine the structure and kinematics of its broad-line region
on the one hand
and its central black hole mass on the other hand.

The study of the kinematics in the central broad-line region of AGN
requires the acquisition of high S/N spectra
for analyzing in detail
velocity resolved line profiles. 
Furthermore, a homogeneous set of spectra  has to be obtained over months
with spacings of days to weeks. 
A comparison of the evolution of line profiles with theoretical models 
(e.g.  Welsh and Horne \cite{welsh91})
can give us information on the kinematics in the broad-line region i.e.
whether radial inflow or outflow motions,
turbulent/chaotic velocity fields, 
or Keplerian orbits are dominant.

We published
first results of a variability campaign of Mrk\,110
with the 9.2m Hobby-Eberly-Telescope in
Kollatschny et al. \cite{kollatschny01}
(hereafter called Paper\,I).
We verified the extent and stratification of the BLR.
Results on the velocity field based on H$\beta$ variations have been
published in 
Kollatschny \& Bischoff \cite{kollatschny02}
(hereafter called Paper\,II)
 indicating
that the broad-line region is connected with a central accretion disk.

Theoretical models of accretion disk outflow scenarios in AGN have been
published by different authors. 
They investigated radiatively-driven wind models (e.g. Murray \& Chiang
 \cite{murray97}, Proga et al. \cite{proga00})
and/or magnetically-driven disk outflow models
(Blandford \& Payne \cite{blandford82}, Bottorff et al.
 \cite{bottorff97},  Emmering et al. \cite{emmering92},
 K\"{o}nigl \& Kartje \cite{koenigl94}).

There are other pieces of evidence from multi-frequency
and/or  spectro-polarimetric observations in the literature
 that the BLR
is connected with an accretion disk
(Elvis \cite{elvis00}, Cohen \& Martel \cite{cohen02}).

In this paper
 we present new velocity delay maps of several Balmer and Helium emission
lines in the spectrum of Mrk\,110 to verify
the accretion disk scenario in the BLR of this galaxy.

%------------------------------------------------------------------------------
%
\section{Observations and data reduction}
The observations of this monitoring campaign
 were carried out at
the 9.2m Hobby-Eberly Telescope (HET) at McDonald Observatory.
We took 26 spectra 
between 1999 November 13 (JD 2,451,495) 
and 2000 May 14 (JD 2,451,678) with a median interval of  3 days.
A log of the observations is presented in Paper\,I.

We obtained all our spectra under
identical conditions with the Low Resolution Spectrograph (LRS) located at
the prime focus of the HET. 
A Ford Aerospace CCD (3072x1024) with 15 $\mu$m pixel
was used throughout our monitoring run.
The spectra cover a wavelength range from
4200\AA\ to 6900\AA\ with a resolving power of 650 at 5000\AA\,. 
% (=450 km/s res
Exposure times were 10 to 20 minutes and yielded
a S/N $>$ 100 per pixel in the continuum in most cases.

We reduced the spectra in a homogeneous way
with IRAF reduction packages including bias subtraction,
cosmic ray correction, flat-field correction,
wavelength calibration, night sky subtraction and flux calibration.
Great care was taken to obtain very good intensity  and
wavelength calibration by using the spatially unresolved forbidden emission
lines in the spectra as additional internal calibrators.

We generated first a mean spectrum of our variability
campaign with very high S/N. Afterwards
we calculated difference spectra of all epochs with respect to this 
mean spectrum. We adapted our spectra
by minimizing the residuals of the constant narrow emission lines
in the difference spectra. The primary lines for the internal calibration
were the [OIII]$\lambda$5007,$\lambda$4959 lines.
Thus we corrected for small spectral shifts, for minor differences in the
spectral resolution caused by different seeing conditions and for small scaling
factors.
We used the narrow components of the Balmer and Helium lines
as well as the [OI]$\lambda$6300,$\lambda$6363, 
[SII]$\lambda$6717,$\lambda$6731,
and [NII]$\lambda$6548,$\lambda$6584 emission lines 
as secondary calibrators
to secure the calibration over the whole spectral range. 
In this way we achieved relative line fluxes with an
accuracy of better than 1\% in most of the spectra.
The main error sources were the variable broad
HeI$\lambda$5016,$\lambda$4922 emission lines blending the
[OIII]$\lambda$5007,$\lambda$4959 calibration lines.
Different seeing conditions during the observations caused  
slightly different stellar contributions from the underlying host galaxy 
in our spectra. 
This effected the pseudo-continuum we used to subtract from the spectra.
Further sources of error are the slopes of the sensitivity functions
we derived from our standard star spectra. At the edges of the spectrum 
one can get varieties of 1\% to 3\% in the sensitivity function
by fitting the observed bandpasses. This impreciseness influenced 
especially the H$\alpha$ and H$\gamma$ 
line profiles. Furthermore, the H$\gamma$ line
is heavily blended with the [OIII]$\lambda$4363 line and the 
H$\alpha$ line is heavily contaminated by the variable atmospheric
absorption.
The errors in the intensities
 of these strong
Balmer line are therefore comparable to those of the weaker Helium
lines.

Further details of the observations and reduction procedure are published in
Papers\,I and II.
\section{Results}
We concentrated on integrated line intensity variations of the broad emission
lines
as well as on line profile variations in the H$\beta$ line
in our Papers\,I and II. Here we present results on line profile variations
in several  Balmer and Helium lines. These line profiles must have a
large S/N ratio for doing this study.
\subsection{Integrated line intensity variations
and rms line profiles}
We derived mean and rms line profiles from all the spectra obtained at our
HET variability campaign.
The rms spectra give us the variable part of the emission lines profiles we are
investigating here.
The constant narrow lines in the galaxy spectra
 cancel out in the rms spectra.
The normalized rms line profiles
of the strongest Balmer (H$\alpha$, H$\beta$)
and Helium emission lines (HeII$\lambda 4686$ , HeI$\lambda 5876$ ) 
are presented in Fig.~1 in velocity space.
\begin{figure}
\includegraphics[bb=40 60 400 700,width=55mm,height=85mm,angle=270]{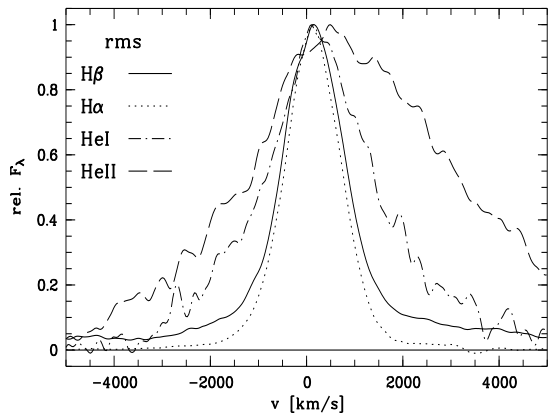}
\caption{Normalized Balmer and Helium rms line profiles in velocity space.}
\end{figure}
We did not consider the H$\gamma$ line in this figure because its 
rms line profile 
is still contaminated by the strong [OIII]$\lambda$4363 line blend.
The big differences in the widths of these lines are obvious.
 The Helium lines and
especially the HeII line are much broader than the Balmer lines.
We published in Paper I (Fig.7)
 the entire rms spectrum of Mrk\,110.
One can see that structures in the broad He line wings are real 
and not caused by noise
by comparing them with the S/N ratio in the continuum.
The measured rms line widths (FWHM) of these profiles are listed in Table\,1.

In Paper\,I we derived
 the mean distance $\tau_{cent}$
 of these line emitting regions 
from the central ionizing source. This was done by
calculating the cross-correlation function
of the integrated line light curves with
the variable ionizing continuum light curve. These results are given
in Table\,1 as well for completeness.
\begin{table}
\caption{Rms line widths (FWHM) of the strongest emission lines,
their cross-correlation lags with respect to continuum variations, and the
derived central black
hole masses (see text for details).}
\begin{tabular}{lcccc}
\hline 
\noalign{\smallskip}
Line & FWHM(rms) & $\tau_{cent}$ & $M$ \\
     & [km s$^{-1}$] & [days] & [$10^7 M_{\odot}$]\\
(1) & (2) & (3) & (4)\\ 
\noalign{\smallskip}
\hline
\noalign{\smallskip}
HeII$\lambda 4686$~ & 4444. $\pm$ 200 & $3.5^{+2.}_{-2.}$ & $2.25^{+1.63}_{-0.45}$\\
HeI$\lambda 5876$~  & 2404. $\pm$ 100 & $10.8^{+4.}_{-4.}$ & $1.81^{+1.36}_{-0.33}$\\
H$\beta$            & 1515. $\pm$ 100 & $23.5^{+4.}_{-4.}$ &  $1.63^{+0.33}_{-0.31}$\\
H$\alpha$           & 1315. $\pm$ 100 & $32.5^{+4.}_{-4.}$ & $1.64^{+0.33}_{-0.35}$ \\
\noalign{\smallskip}
\hline 
\end{tabular}
\end{table}
In Fig.~2 we plot the derived time lags 
of the Balmer and Helium emission lines 
as a function of their FWHM of the rms line profiles. 
These time lags can be interpreted as the light-travel time
across the emission region for the different broad emission lines.
There is a clear correlation between line width and time lag.

We interprete the light-travel time
as the characteristic distance $R$ of the line emitting region and
the FWHM of the rms emission line width 
as the characteristic velocity $v$ of the line emitting clouds.
We estimate the central masses of the central black hole
under the assumption that the gas
dynamics are dominated by a central massive object (see Paper\,I).
Additionally, we calculated the relation between radius and velocity
for different black hole masses.
The dotted and dashed lines shown in Fig.~2 correspond to
virial masses of 
0.8, 1.5, 1.8, 2.2, and 2.9 $\cdot10^{7} M_{\odot}$ (from bottom to top).
\begin{figure}
 \includegraphics[bb=40 45 380 700,width=70mm,angle=270]{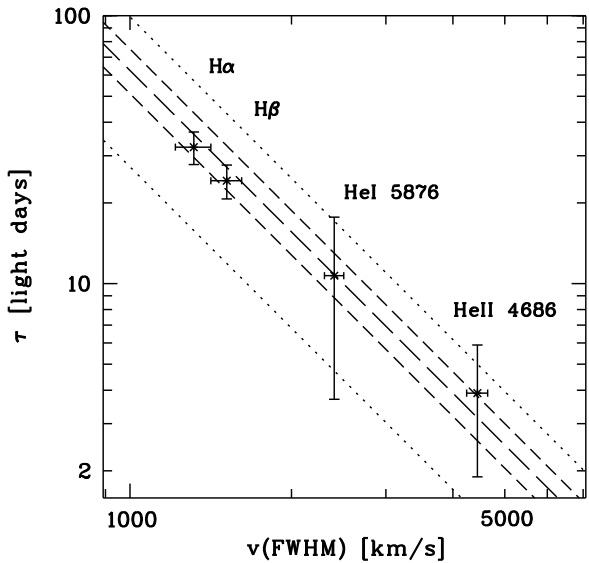}
\vspace*{5mm}
\caption{The distance of the Balmer and Helium emitting line regions
from the central ionizing source in Mrk\,110
as a function of the FWHM in their rms line profiles.
The dotted and dashed lines are the results from our model calculations
for central masses of 
0.8, 1.5, 1.8, 2.2, and 2.9 $\cdot10^{7} M_{\odot}$ (from bottom to top).}
\end{figure}
It is evident that the black hole mass of 1.8 $\cdot10^{7} M_{\odot}$
we derived in Paper I from the observed data  of the individual
emission lines matches the calculated model relation.
But we have to keep in mind that
there might be additional systematic
uncertainties in the determination of the central black hole mass
due to not considered geometry and/or orientation
effects of the BLR in these models.
\subsection{Light curves of emission line segments}
The light curves of the continuum flux at 5135\AA\ and of the integrated
broad emission line intensities 
of our HET variability campaign are
published in Paper\,I. The 
strongest broad emission lines in our spectra are the
H$\alpha$, H$\beta$, and H$\gamma$
Balmer lines and the
HeII$\lambda$4686 and HeI$\lambda$5876 Helium lines.

In this paper we concentrate on variations
in the emission line profile only.
First results of the H$\beta$ profile variations
have been published in Paper II. 
We measured the light curves of all subsequent
velocity segments ($\Delta$$v$ = 400 km/s width)
from $v$ = -5000 until +5000 km/s
in the strongest Balmer and Helium lines.
The intensity of
the central line segment is integrated from
$v$ = -200 until +200 km/s.
The light curves of
the line center as well as of blue and red line wing
 segments at $v$ = $\mp$ 600, 1200, 2000 km/s  ($\Delta$$v$ = 400 km/s)
for HeII$\lambda$4686, HeI$\lambda$5876, H$\gamma$, and H$\alpha$
are shown in Figs.~3 to 6
in addition to the continuum light curve.
In Paper\,II we published corresponding
light curves for H$\beta$. In that case
we used segments with
a velocity binning of $\Delta$$v$ = 200 km/s. 
But these light curves are nearly identical to those resulting with a
$\Delta$$v$ = 400 km/s binning.

The H$\gamma$ line segment light curves are heavily contaminated
by other lines.
The red wing
is heavily blended with the strong [OIII]$\lambda$4363 line
(Paper I, Fig.~1 ) and the blue wing by FeII multiplets.
The red wing of the
redshifted H$\alpha$ line is heavily contaminated by
atmospheric absorption in addition to the blending by the
[NII]$\lambda$6548,6584 lines within
the line profile.  
%
%%
%\begin{figure*}
% \hbox{\includegraphics[bb=40 90 380 700,width=55mm,height=85mm,angle=270]
%{hetcon50lc.ps}\hspace*{7mm}
%       \includegraphics[bb=40 90 380 700,width=55mm,height=85mm,angle=270]
%{m110lcabhbcend4.ps}}
% \hbox{\includegraphics[bb=40 90 380 700,width=55mm,height=85mm,angle=270]
%{m110lcabhbm0600d4.ps}\hspace*{7mm}
%       \includegraphics[bb=40 90 380 700,width=55mm,height=85mm,angle=270]
%{m110lcabhbp0600d4.ps}}
% \hbox{\includegraphics[bb=40 90 380 700,width=55mm,height=85mm,angle=270]
%{m110lcabhbm1200d4.ps}\hspace*{7mm}
%       \includegraphics[bb=40 90 380 700,width=55mm,height=85mm,angle=270]
%{m110lcabhbp1200d4.ps}}
% \hbox{\includegraphics[bb=40 90 380 700,width=55mm,height=85mm,angle=270]
%{m110lcabhbm2000d4.ps}\hspace*{7mm}
%       \includegraphics[bb=40 90 380 700,width=55mm,height=85mm,angle=270]
%{m110lcabhbp2000d4.ps}}
%       \vspace*{5mm}
%  \caption{Light curves of the continuum, of the H$\beta$ line center
% as well as of different blue and red H$\beta$ line wing segments
% ($v$ = $\mp$ 600, 1200, 2000 km/s, $\Delta$$v$ = 400 km/s) derived from our
% HET variability campaign of Mrk\,110.}
%% and 4265\,\AA\
%%
%\end{figure*}
%
\begin{figure*}
 \hbox{\includegraphics[bb=40 90 380 700,width=55mm,height=85mm,angle=270]
{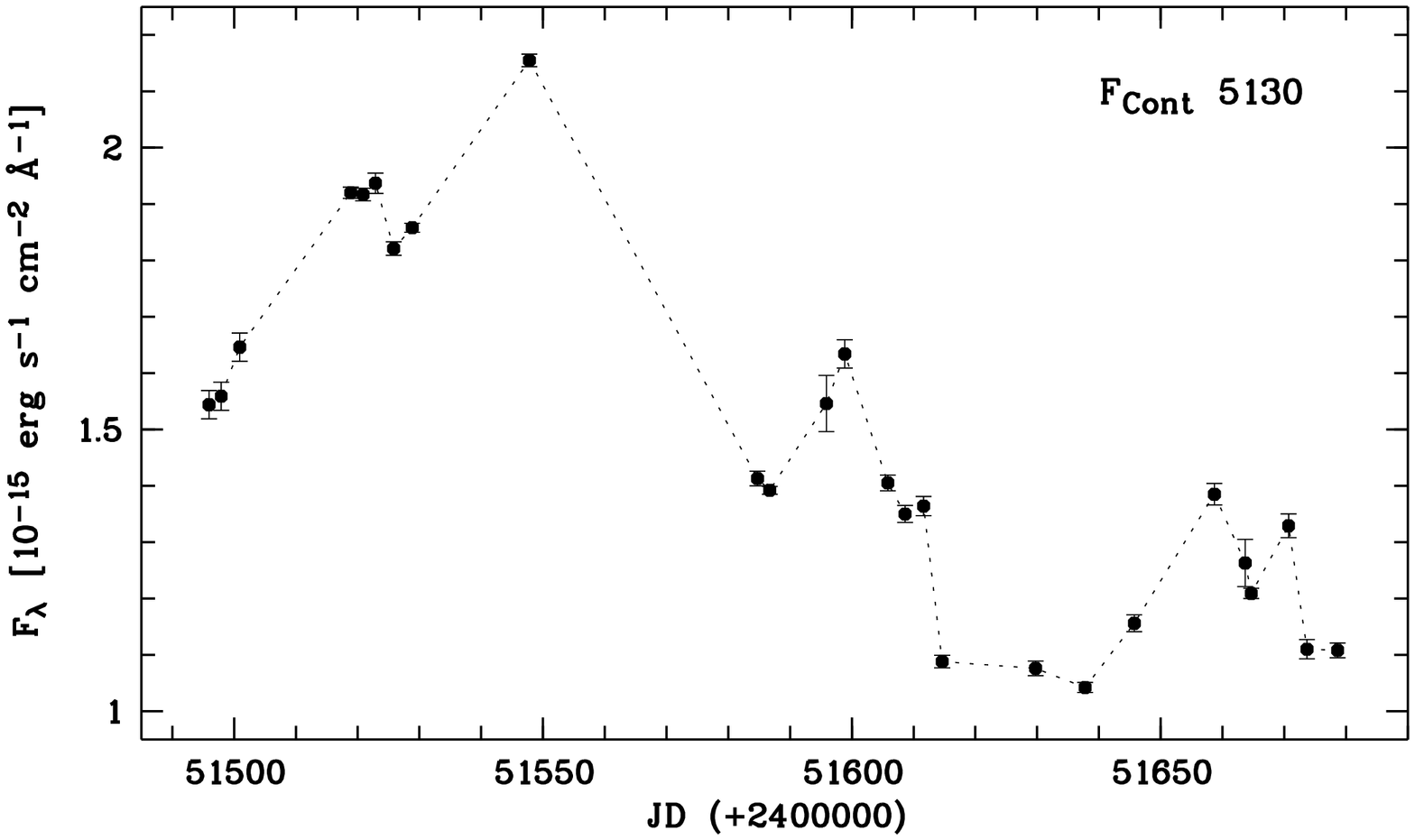}\hspace*{7mm}
       \includegraphics[bb=40 90 380 700,width=55mm,height=85mm,angle=270]
{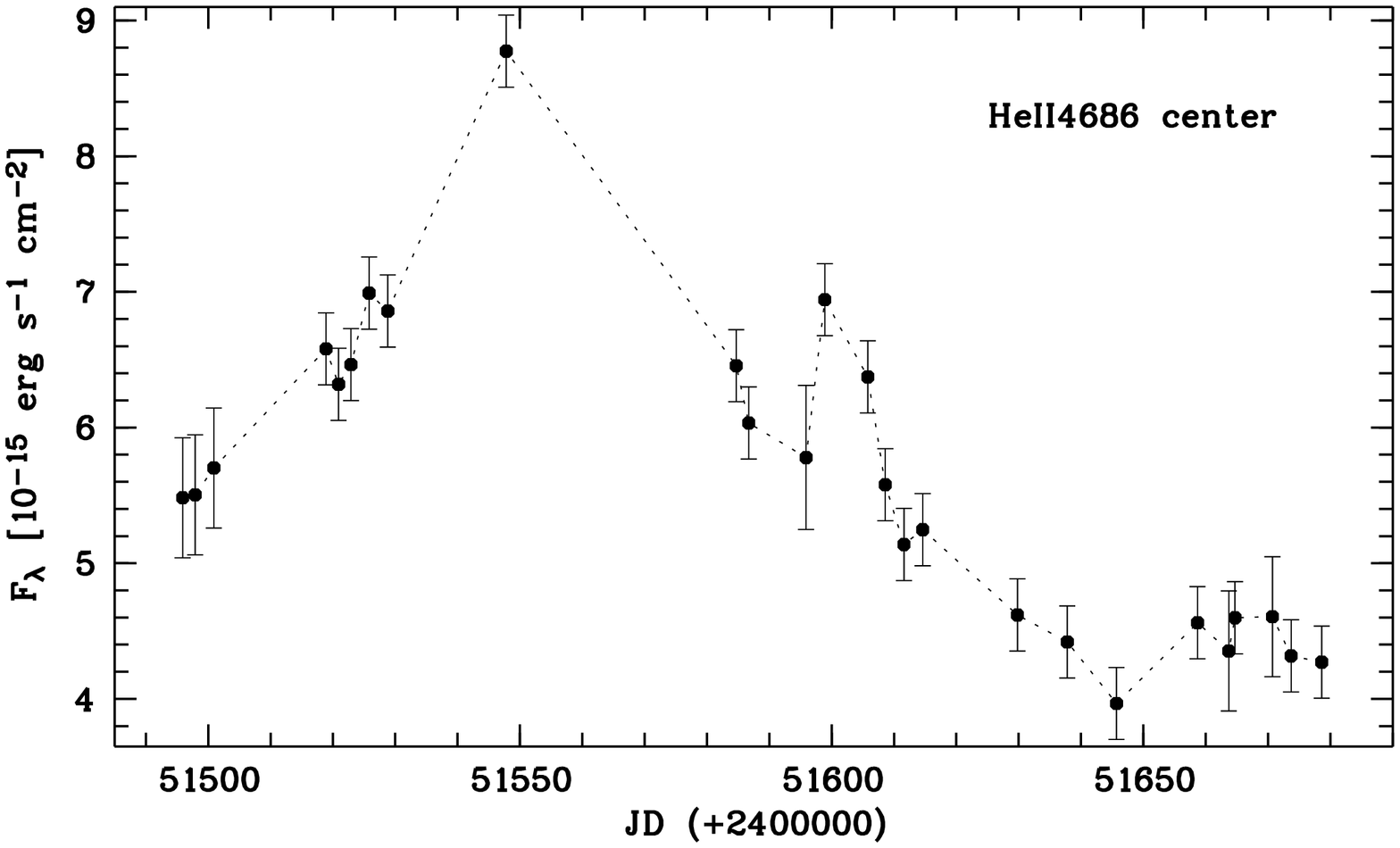}}
 \hbox{\includegraphics[bb=40 90 380 700,width=55mm,height=85mm,angle=270]
{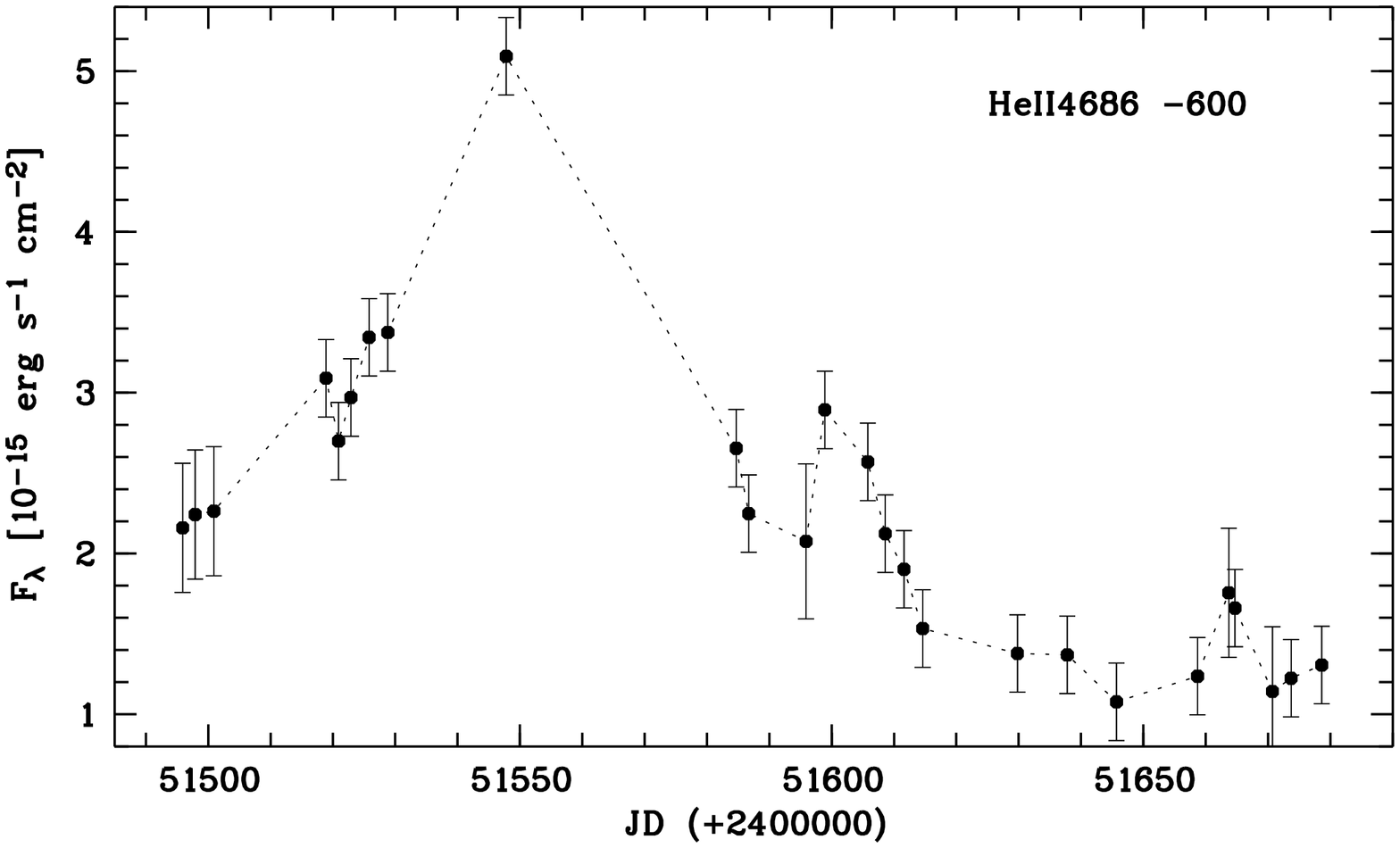}\hspace*{7mm}
       \includegraphics[bb=40 90 380 700,width=55mm,height=85mm,angle=270]
{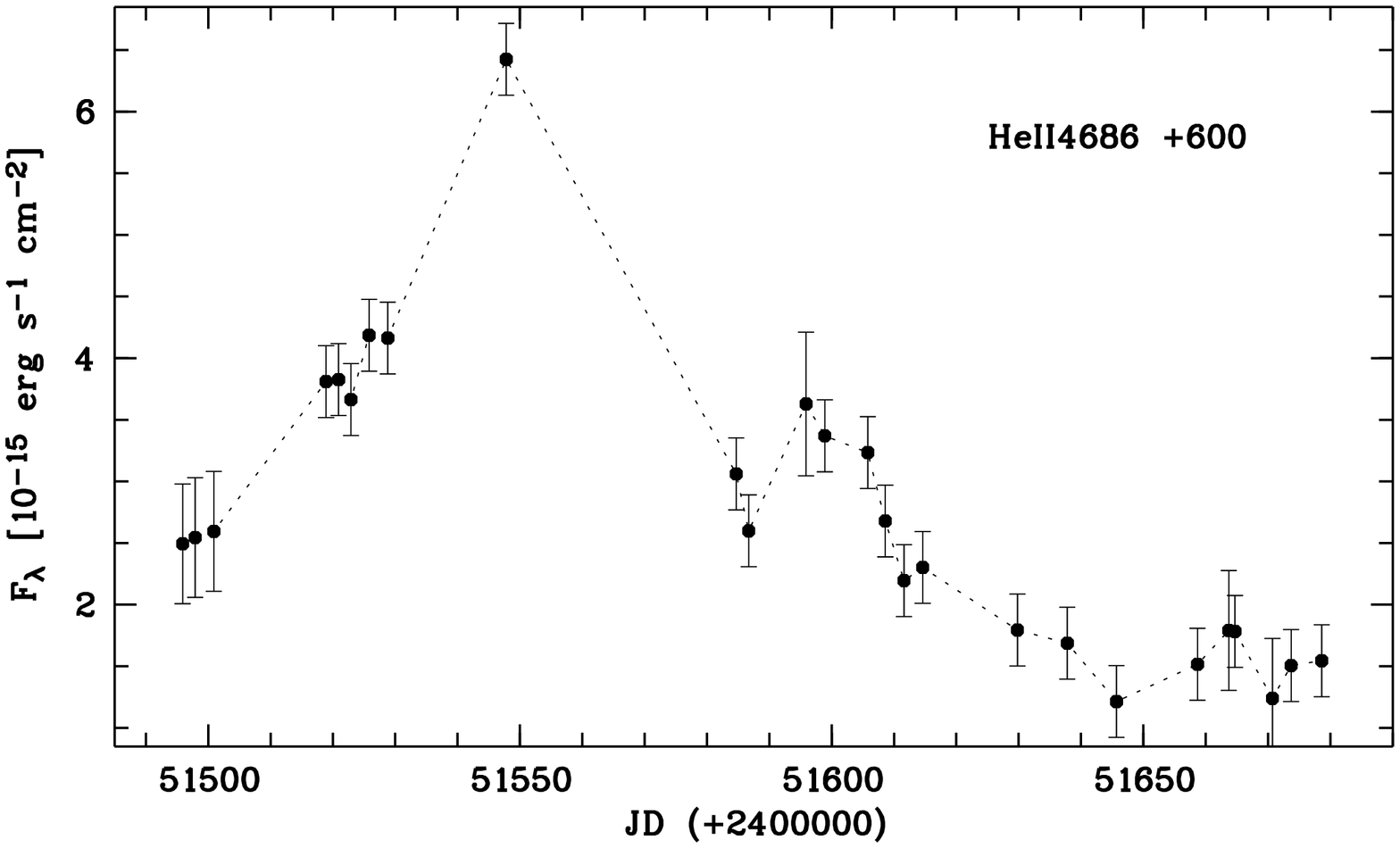}}
 \hbox{\includegraphics[bb=40 90 380 700,width=55mm,height=85mm,angle=270]
{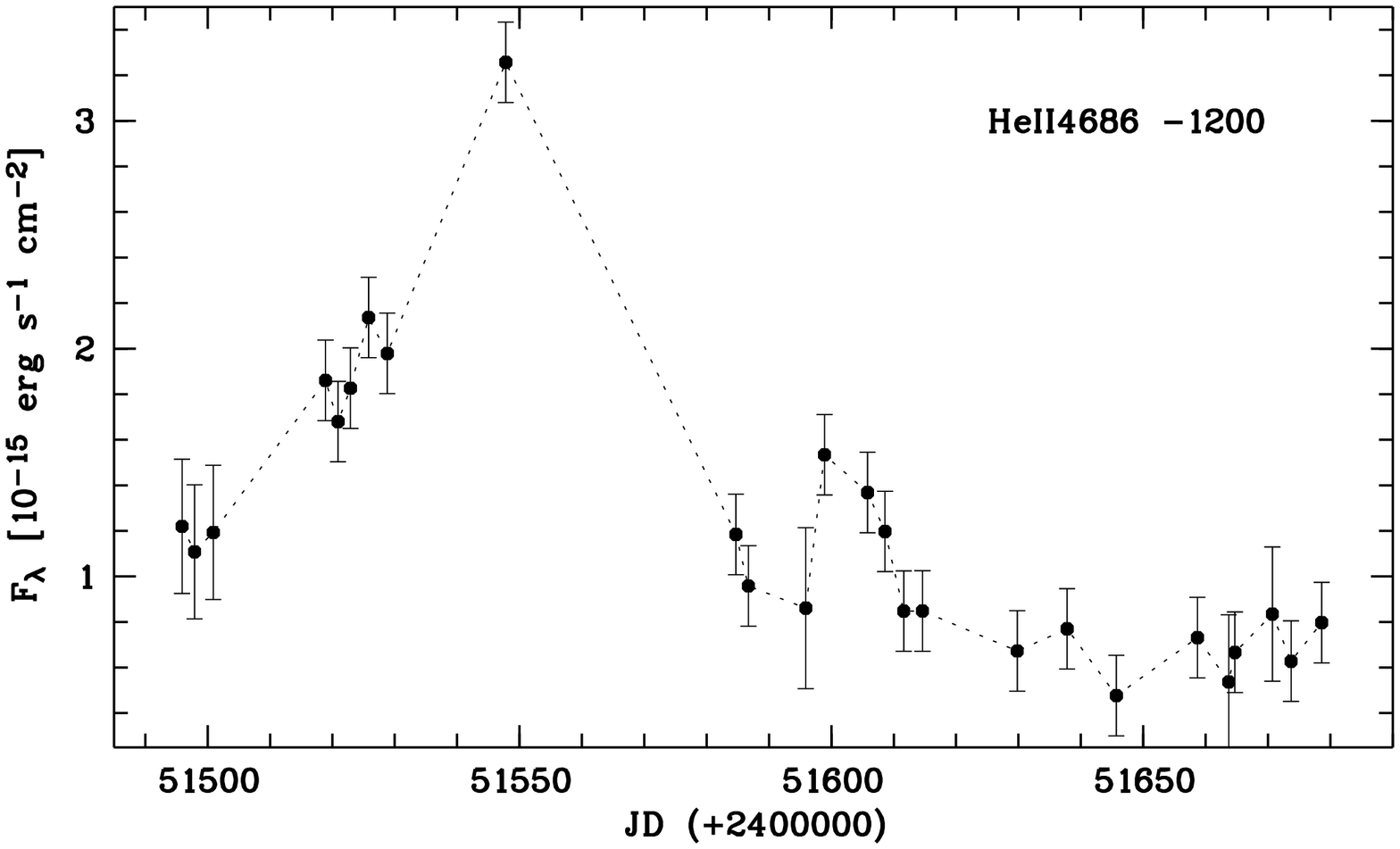}\hspace*{7mm}
       \includegraphics[bb=40 90 380 700,width=55mm,height=85mm,angle=270]
{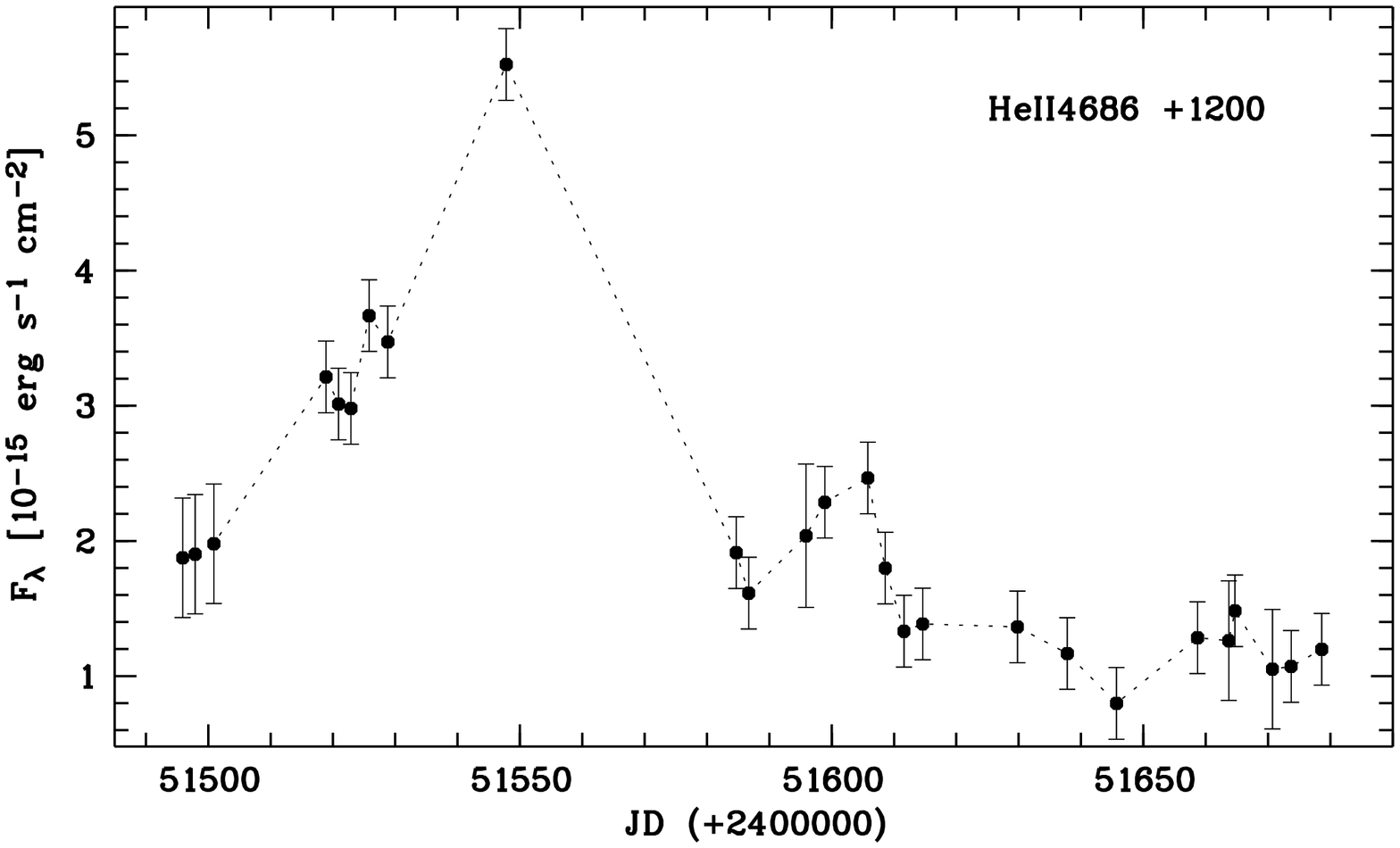}}
 \hbox{\includegraphics[bb=40 90 380 700,width=55mm,height=85mm,angle=270]
{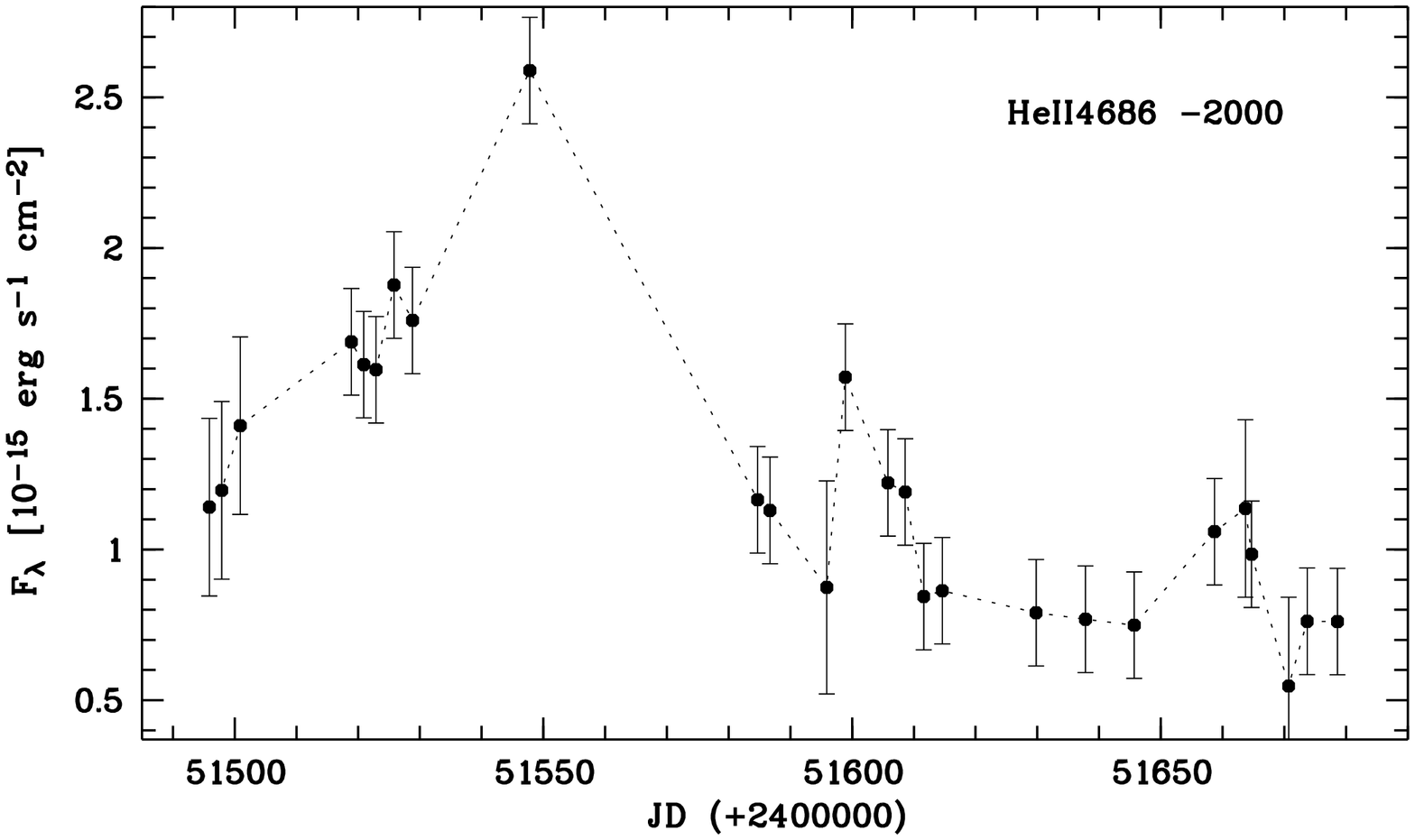}\hspace*{7mm}
       \includegraphics[bb=40 90 380 700,width=55mm,height=85mm,angle=270]
{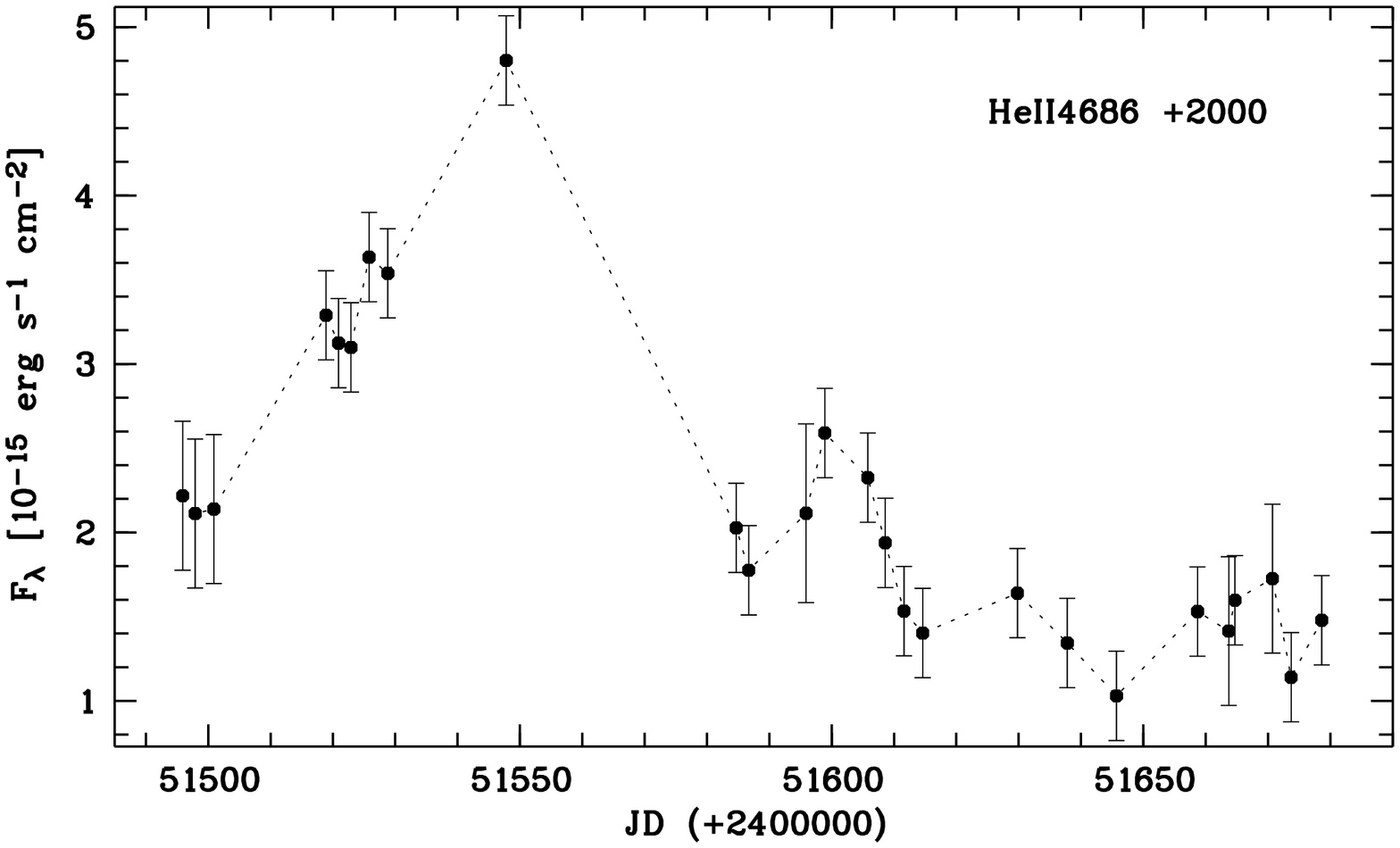}}
       \vspace*{5mm}
  \caption{Light curves of the continuum, of the HeII$\lambda$4686 line center
 as well as of different blue and red HeII$\lambda$4686 line wing segments
 ($v$ = $\mp$ 600, 1200, 2000 km/s, $\Delta$$v$ = 400 km/s) derived from our
 HET variability campaign of Mrk\,110.}
% and 4265\,\AA\
%
\end{figure*}
\begin{figure*}
 \hbox{\includegraphics[bb=40 90 380 700,width=55mm,height=85mm,angle=270]
{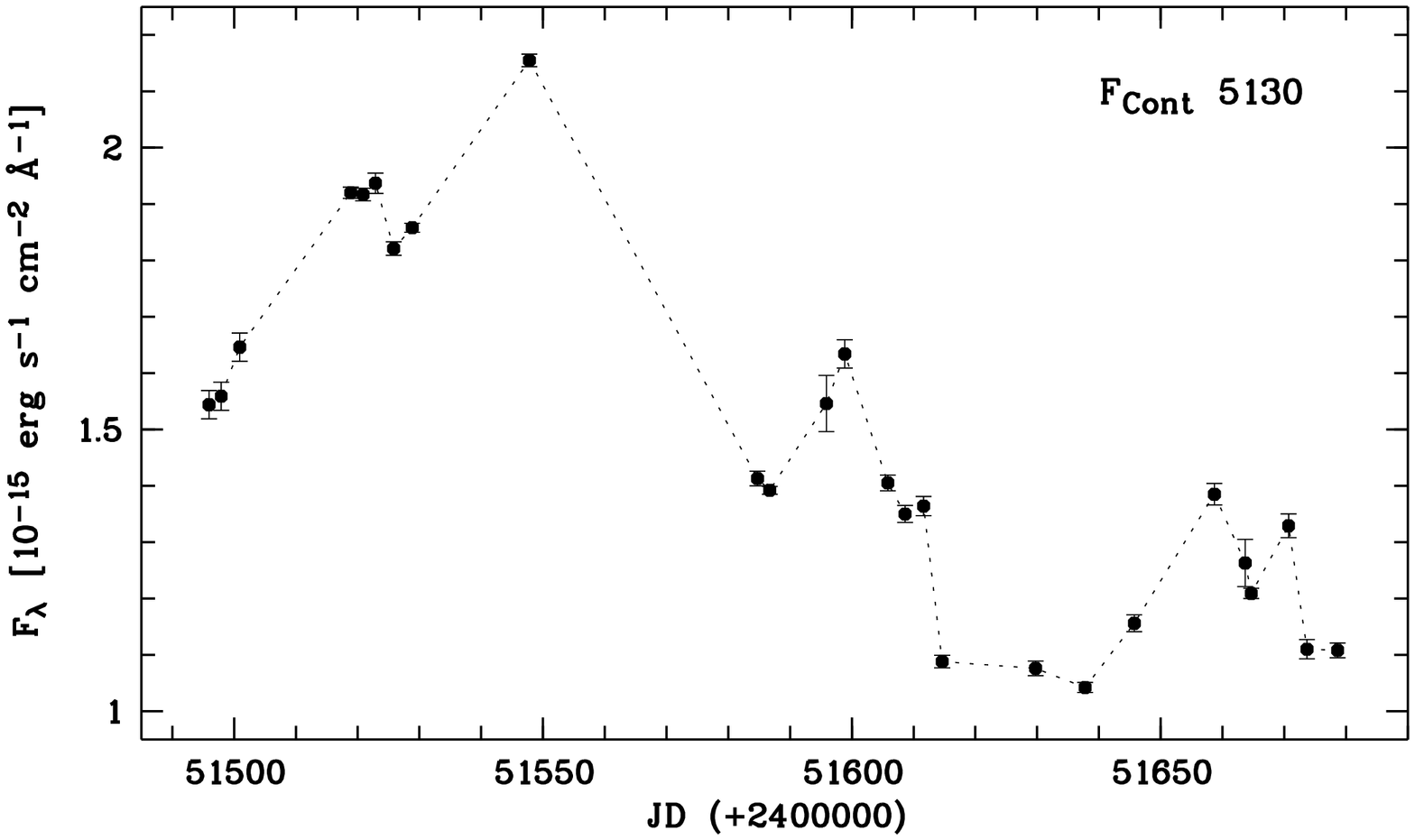}\hspace*{7mm}
       \includegraphics[bb=40 90 380 700,width=55mm,height=85mm,angle=270]
{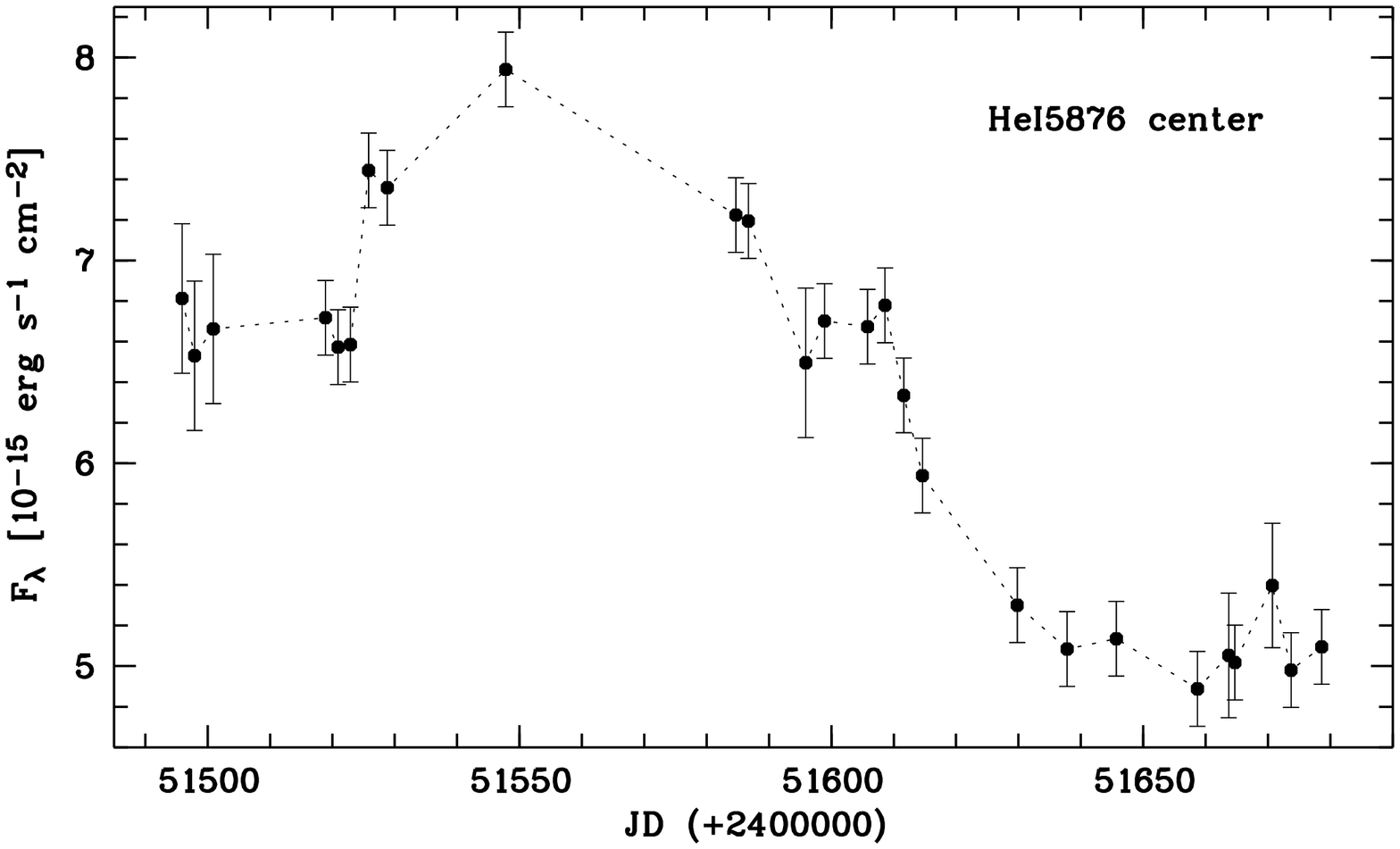}}
 \hbox{\includegraphics[bb=40 90 380 700,width=55mm,height=85mm,angle=270]
{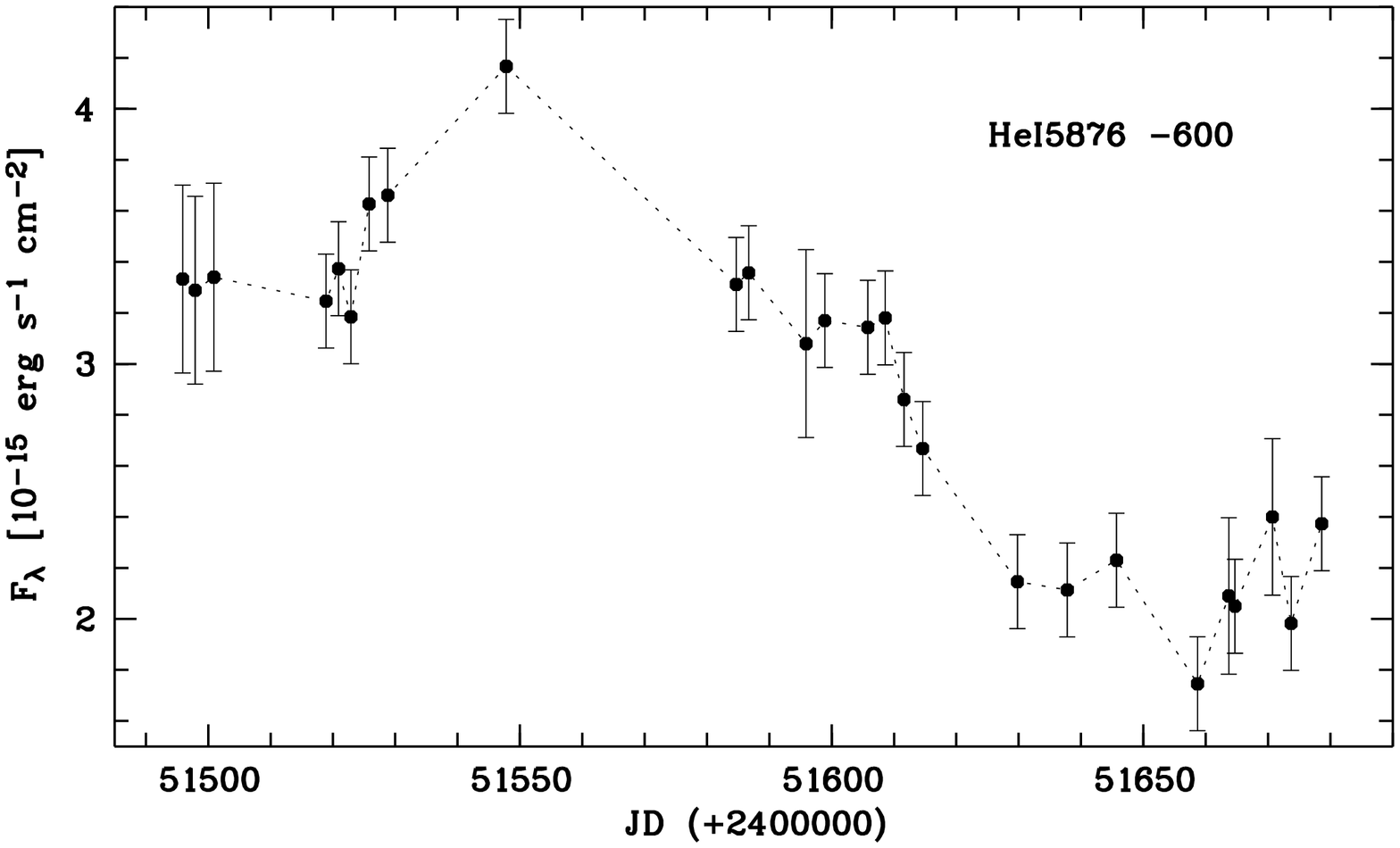}\hspace*{7mm}
       \includegraphics[bb=40 90 380 700,width=55mm,height=85mm,angle=270]
{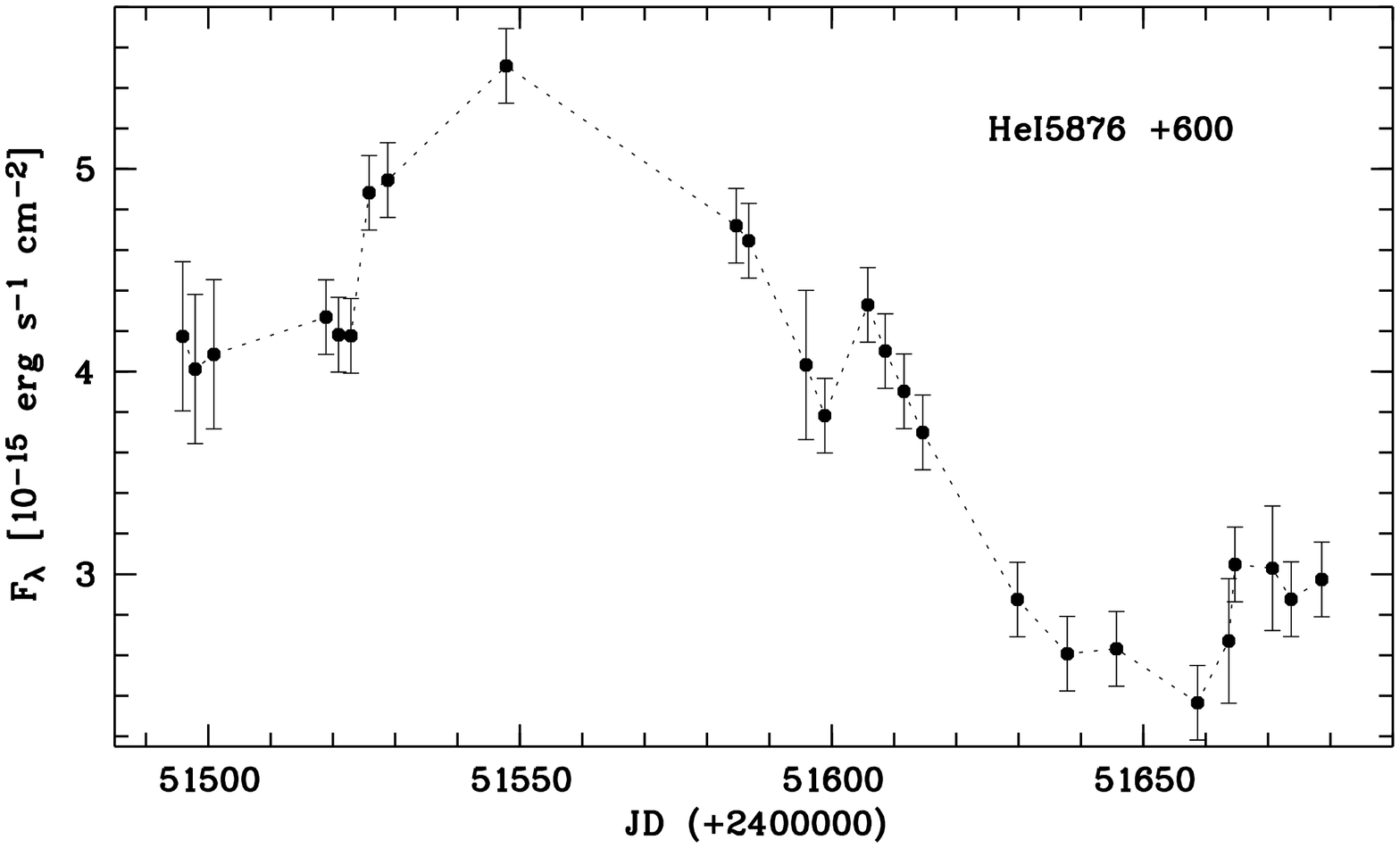}}
 \hbox{\includegraphics[bb=40 90 380 700,width=55mm,height=85mm,angle=270]
{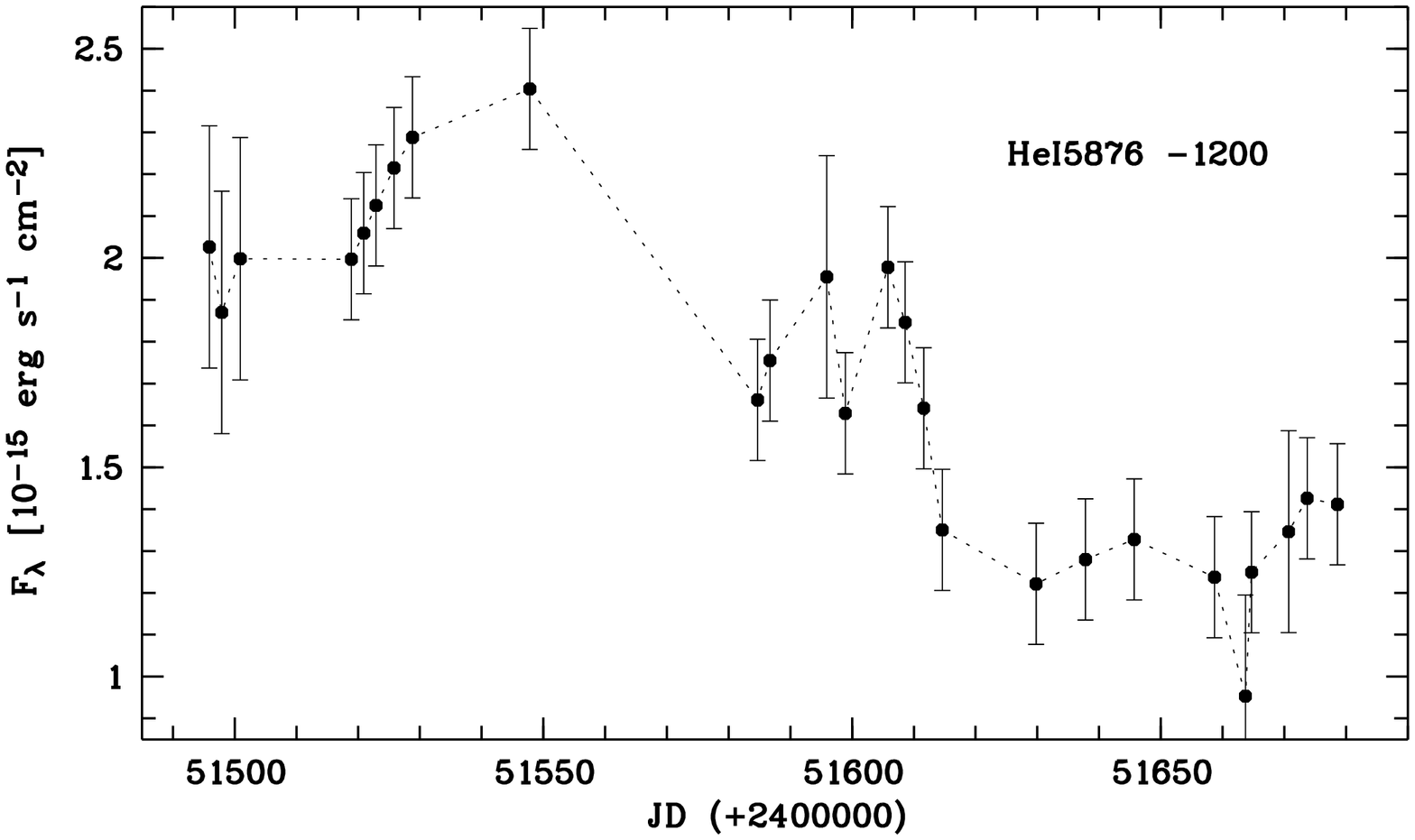}\hspace*{7mm}
       \includegraphics[bb=40 90 380 700,width=55mm,height=85mm,angle=270]
{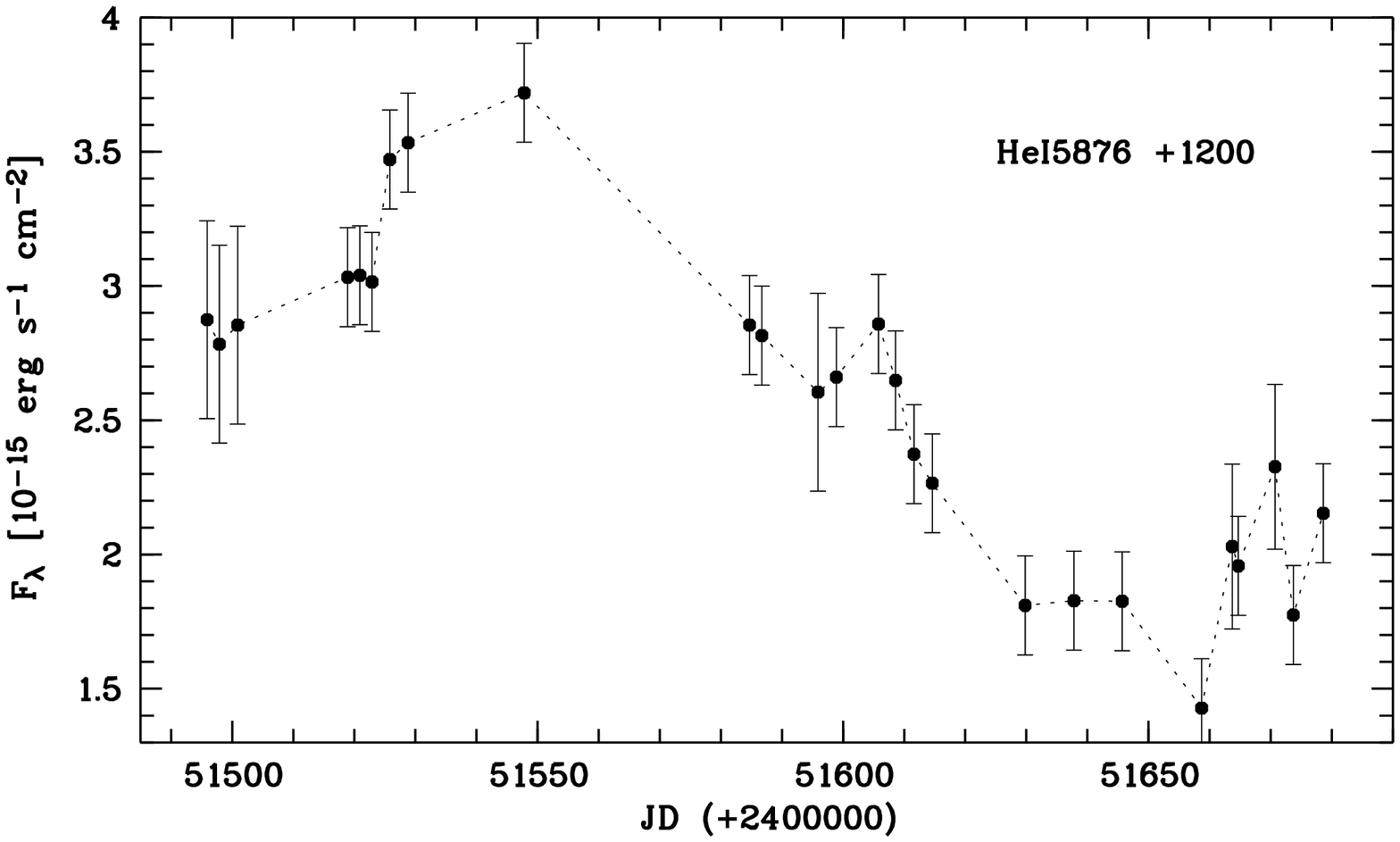}}
 \hbox{\includegraphics[bb=40 90 380 700,width=55mm,height=85mm,angle=270]
{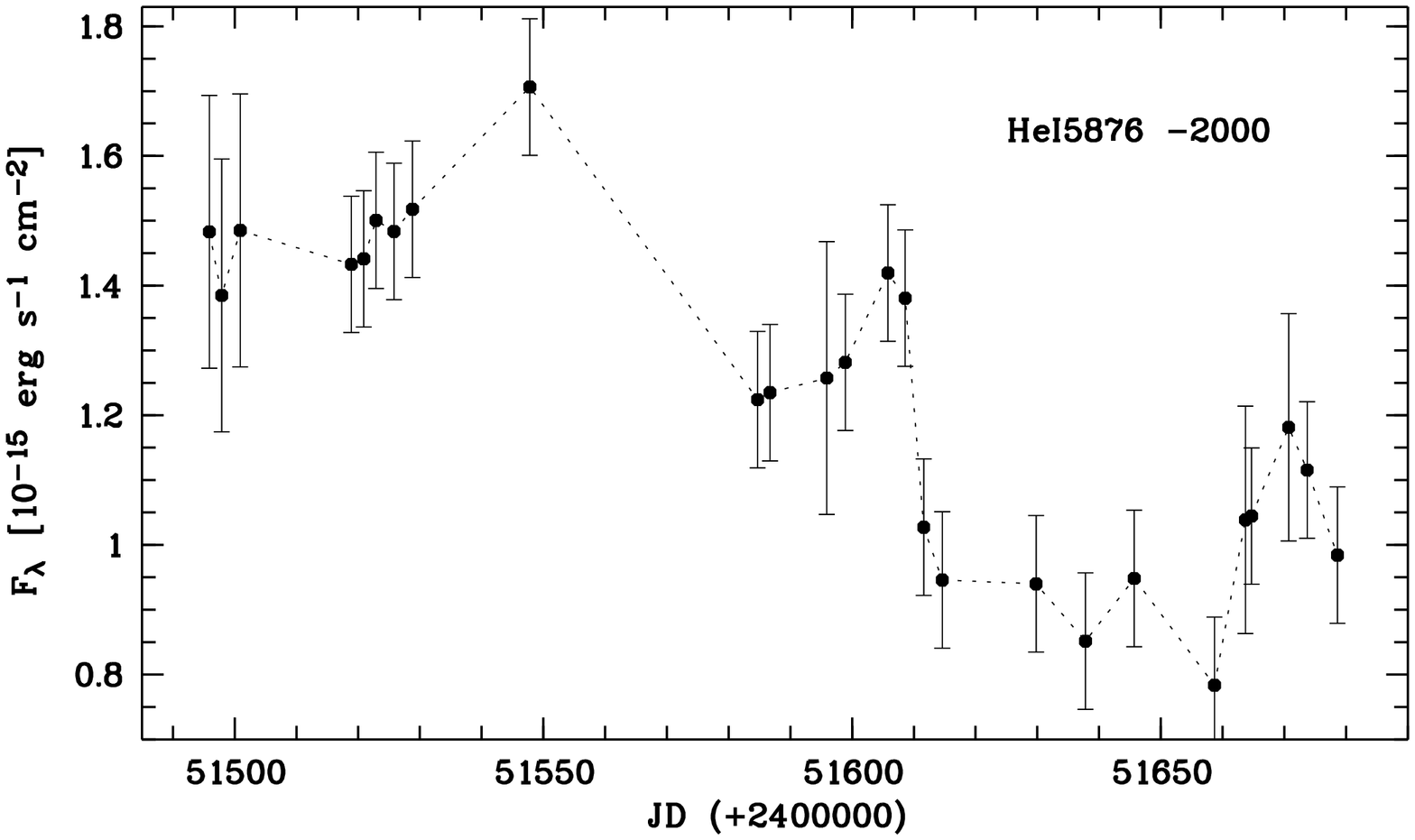}\hspace*{7mm}
       \includegraphics[bb=40 90 380 700,width=55mm,height=85mm,angle=270]
{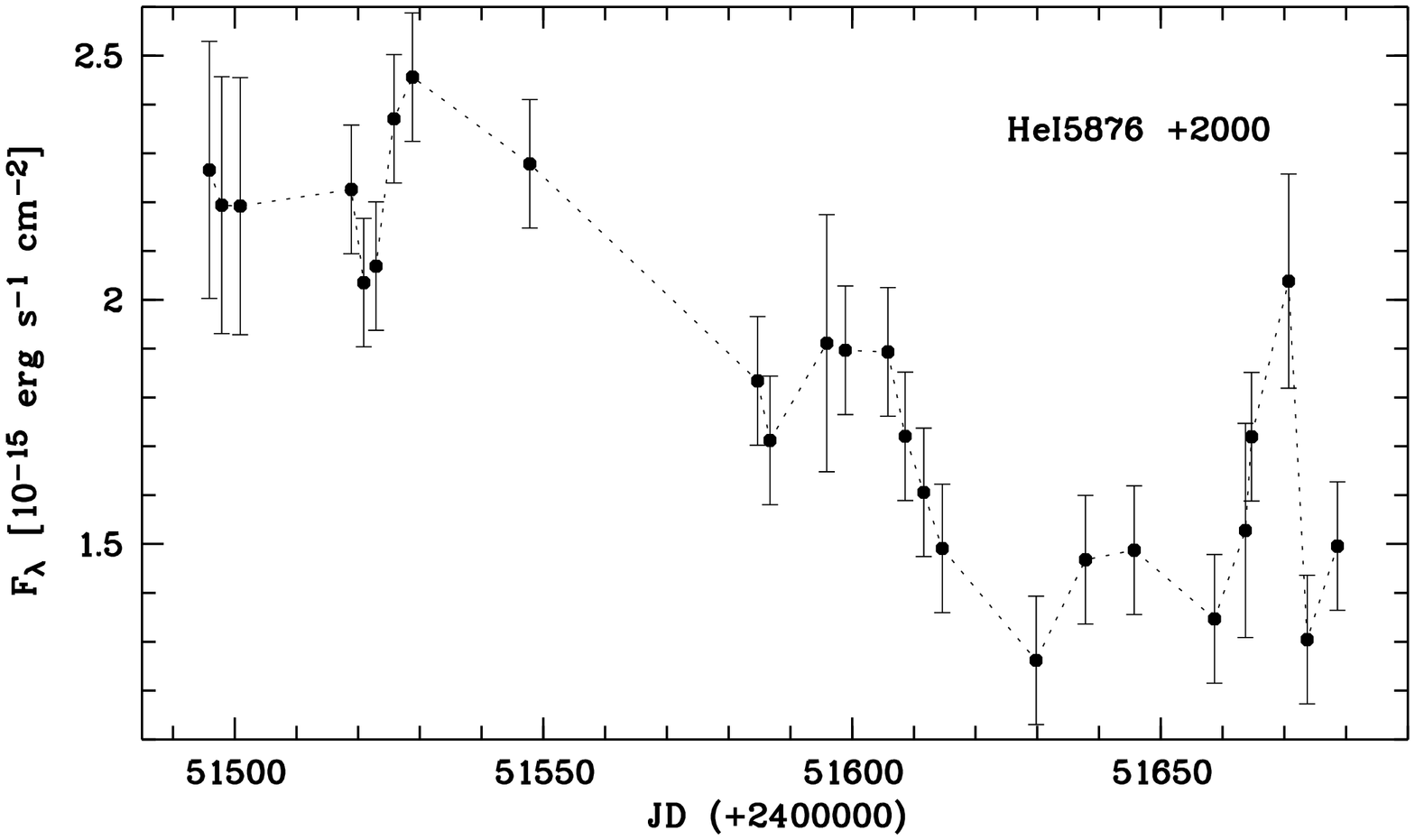}}
       \vspace*{5mm}
  \caption{Light curves of the continuum, of the HeI$\lambda$5876 line center
 as well as of different blue and red  HeI$\lambda$5876 line wing segments
 ($v$ = $\mp$ 600, 1200, 2000 km/s, $\Delta$$v$ = 400 km/s) of Mrk\,110.}
% and 4265\,\AA\
%
\end{figure*}
\begin{figure*}
 \hbox{\includegraphics[bb=40 90 380 700,width=55mm,height=85mm,angle=270]
{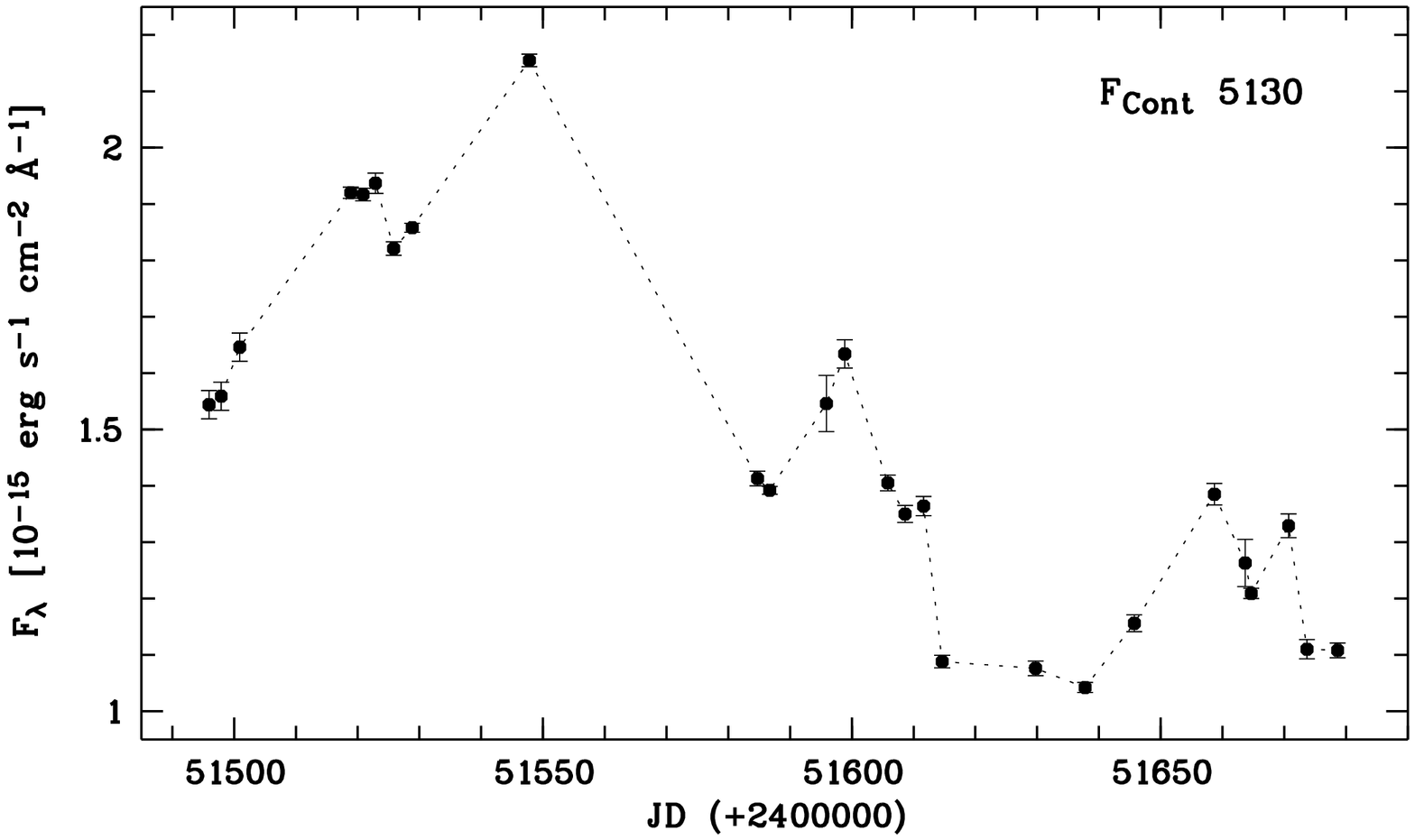}\hspace*{7mm}
       \includegraphics[bb=40 90 380 700,width=55mm,height=85mm,angle=270]
{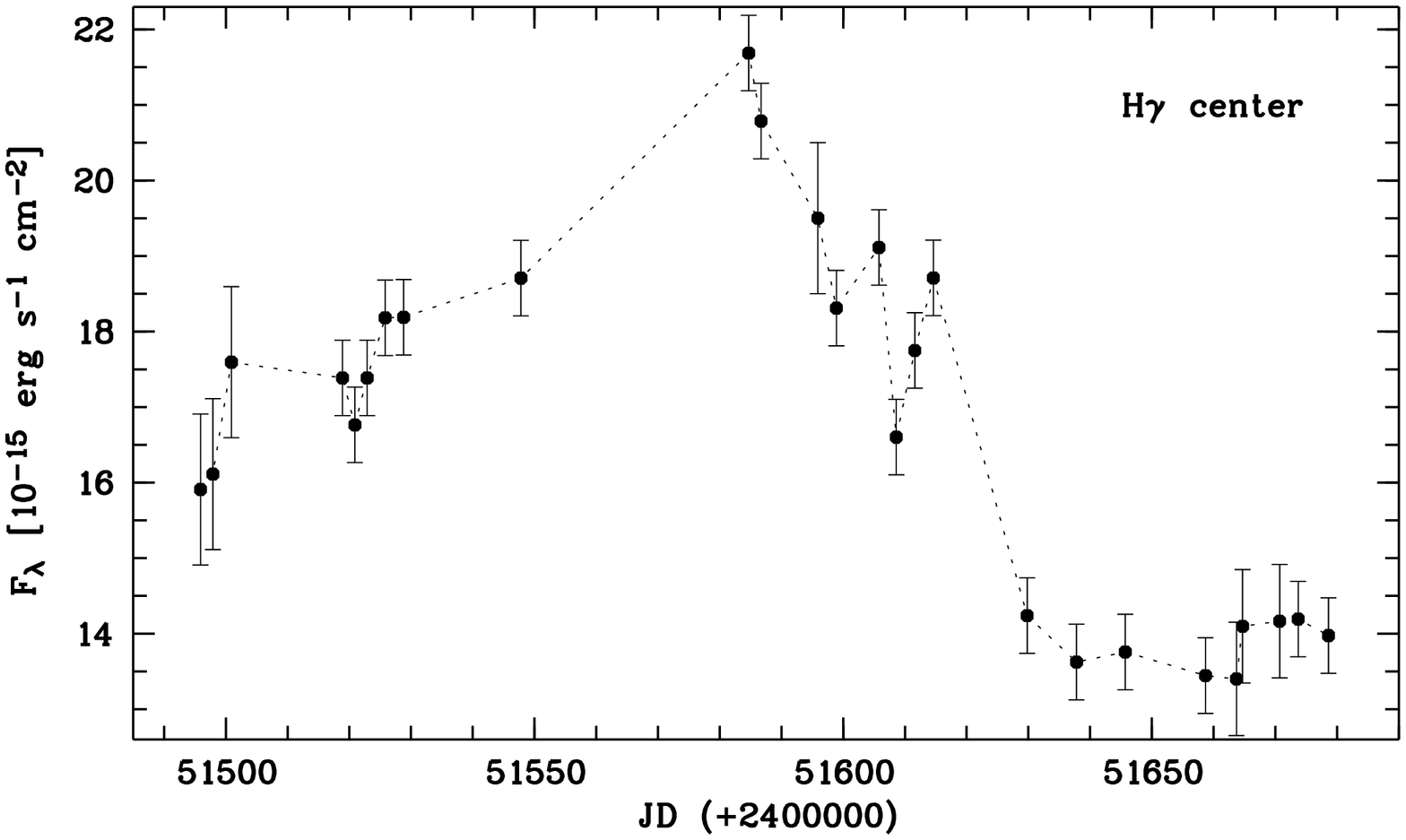}}
 \hbox{\includegraphics[bb=40 90 380 700,width=55mm,height=85mm,angle=270]
{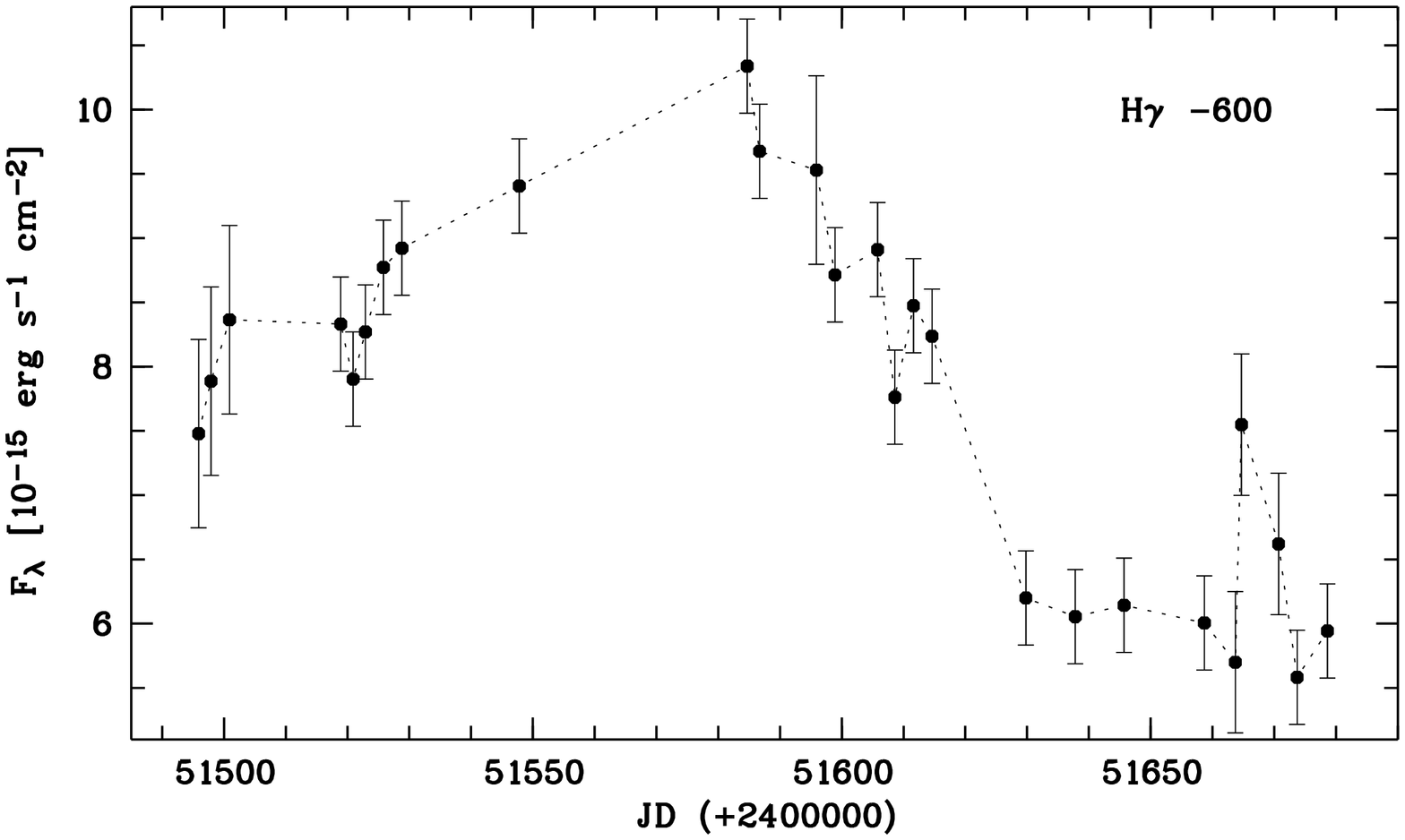}\hspace*{7mm}
       \includegraphics[bb=40 90 380 700,width=55mm,height=85mm,angle=270]
{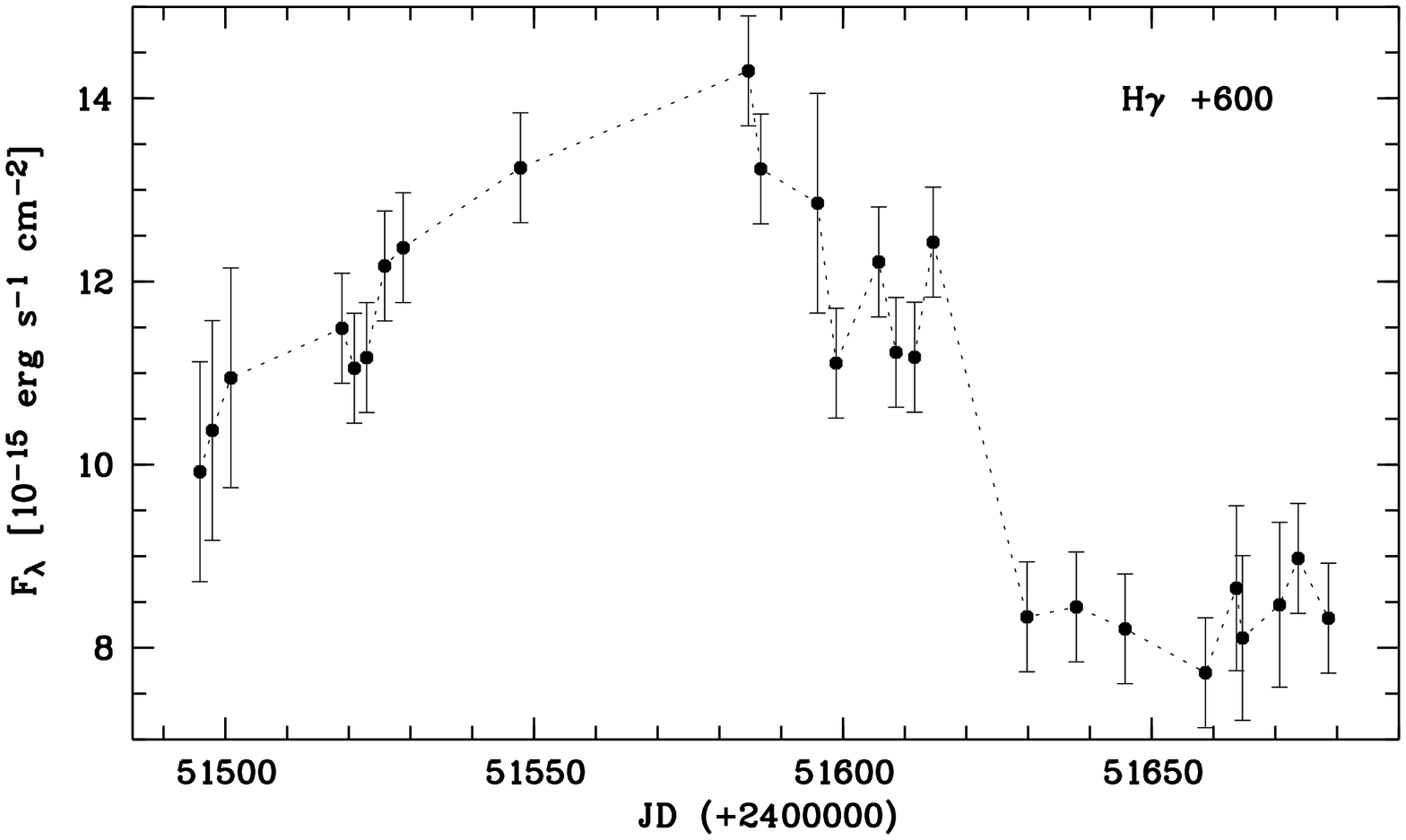}}
 \hbox{\includegraphics[bb=40 90 380 700,width=55mm,height=85mm,angle=270]
{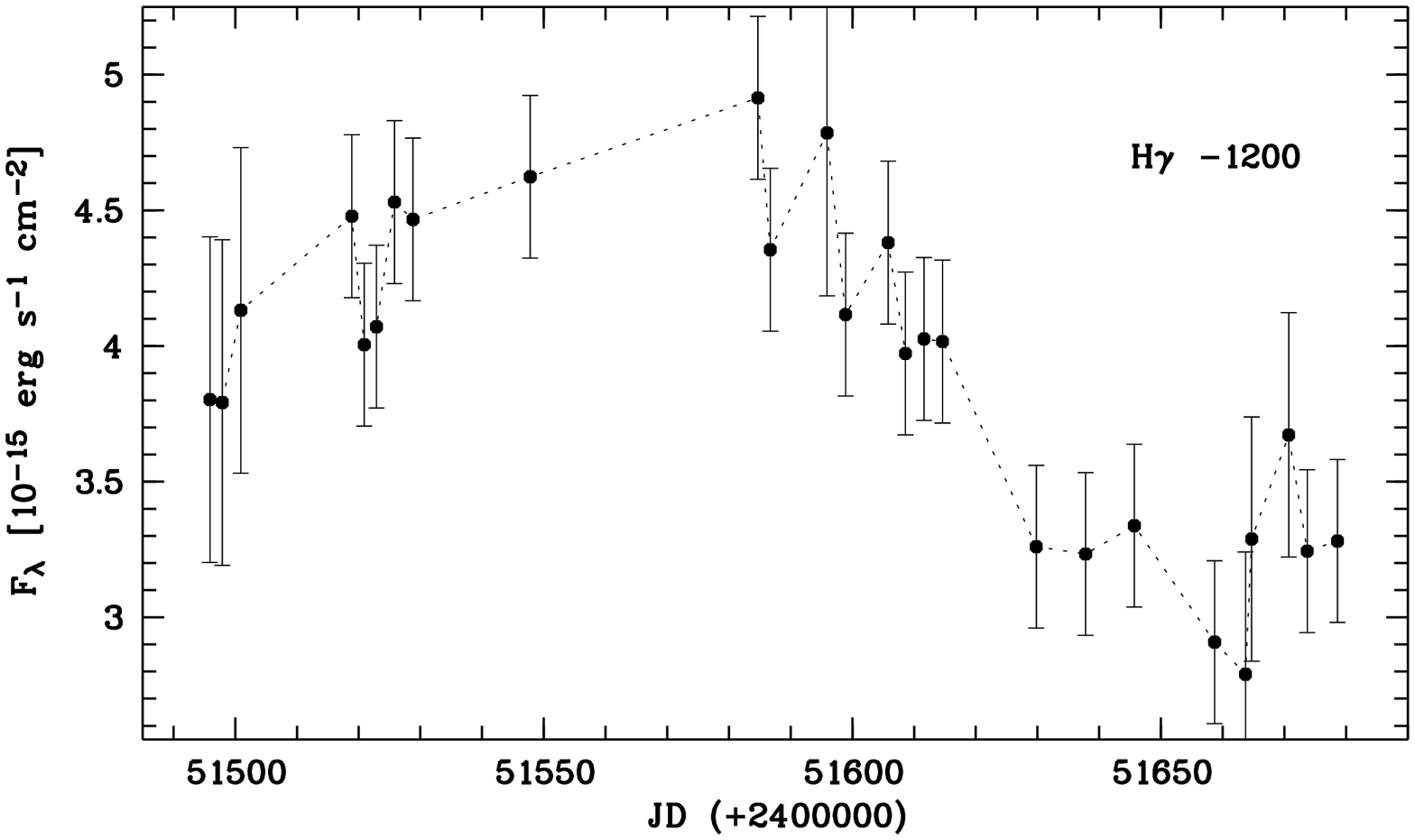}\hspace*{7mm}
       \includegraphics[bb=40 90 380 700,width=55mm,height=85mm,angle=270]
{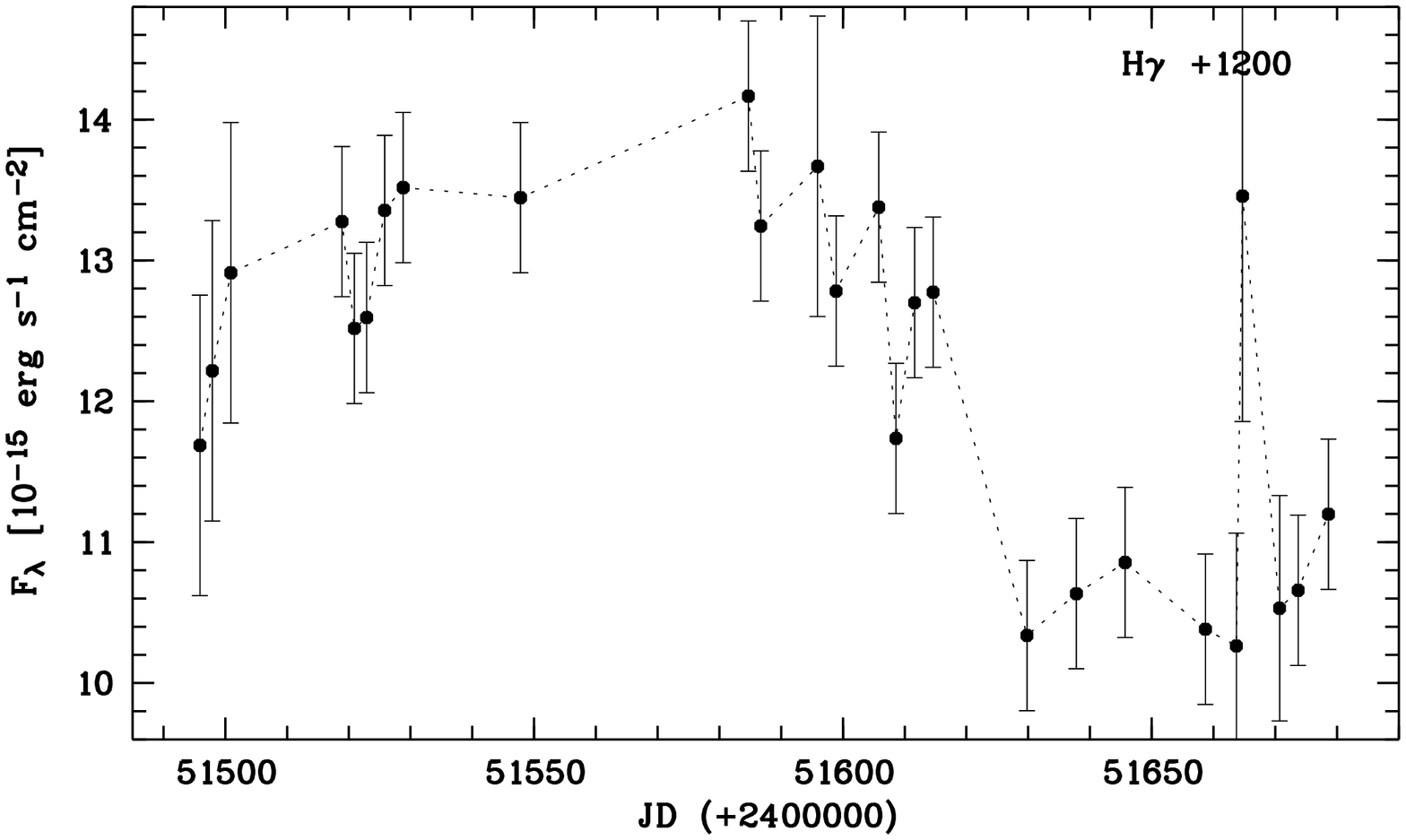}}
 \hbox{\includegraphics[bb=40 90 380 700,width=55mm,height=85mm,angle=270]
{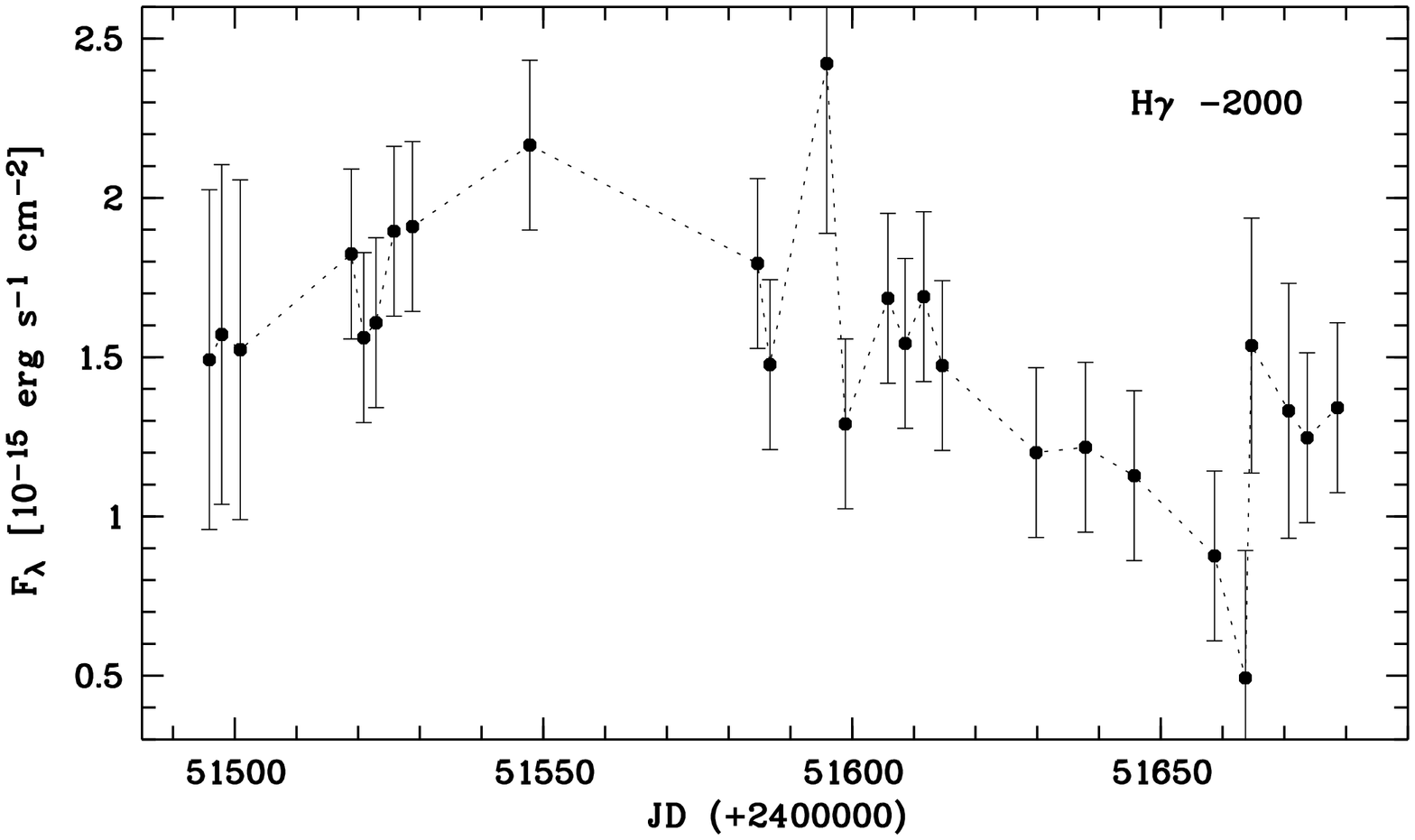}\hspace*{7mm}
       \includegraphics[bb=40 90 380 700,width=55mm,height=85mm,angle=270]
{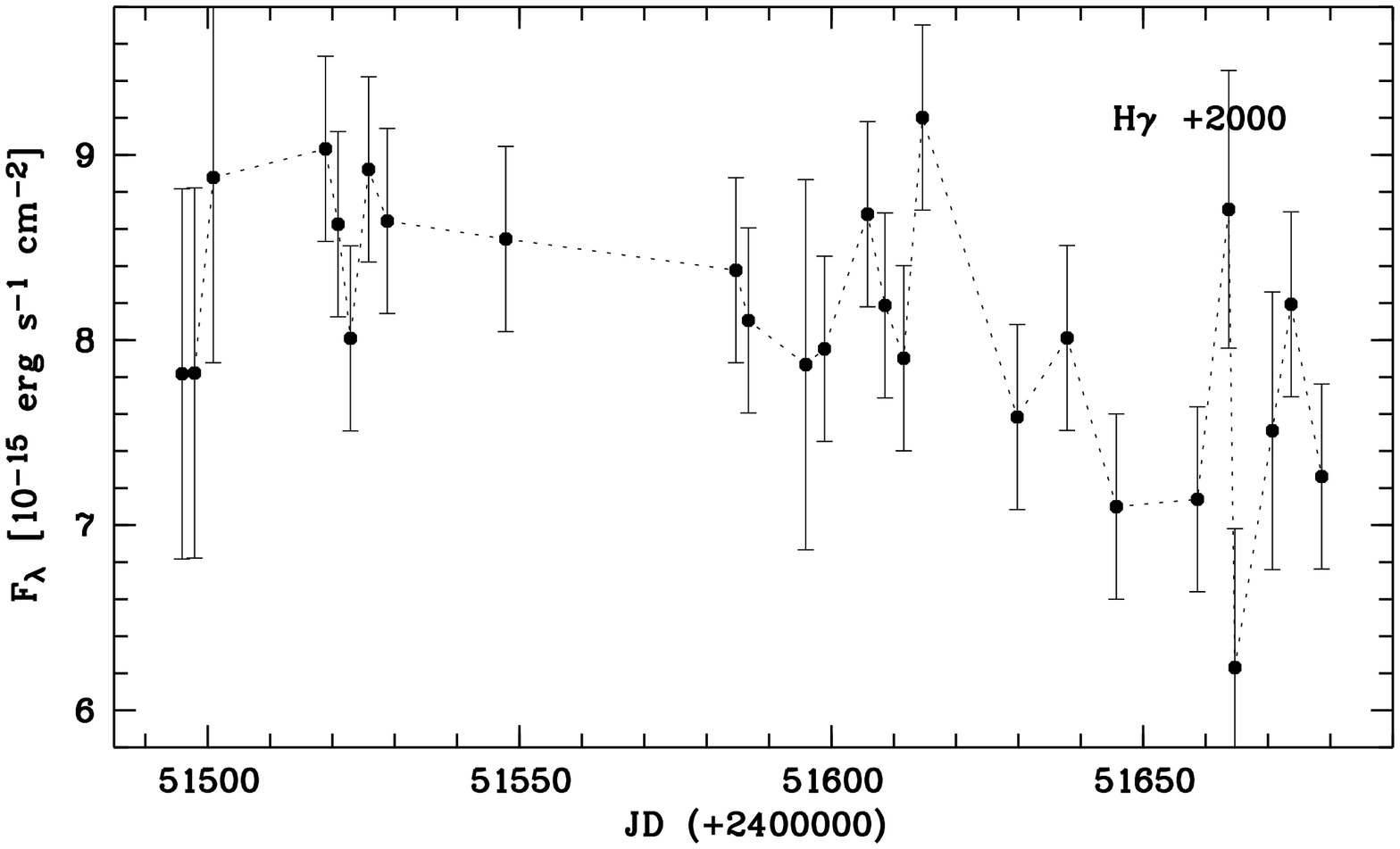}}
       \vspace*{5mm}
  \caption{Light curves of the continuum, of the H$\gamma$ line center
 as well as of different blue and red H$\gamma$ line wing segments
 ($v$ = $\pm$ 600, 1200, 2000 km/s, $\Delta$$v$ = 400 km/s) of Mrk\,110.}
% and 4265\,\AA\
%
\end{figure*}
\begin{figure*}
 \hbox{\includegraphics[bb=40 90 380 700,width=55mm,height=85mm,angle=270]
{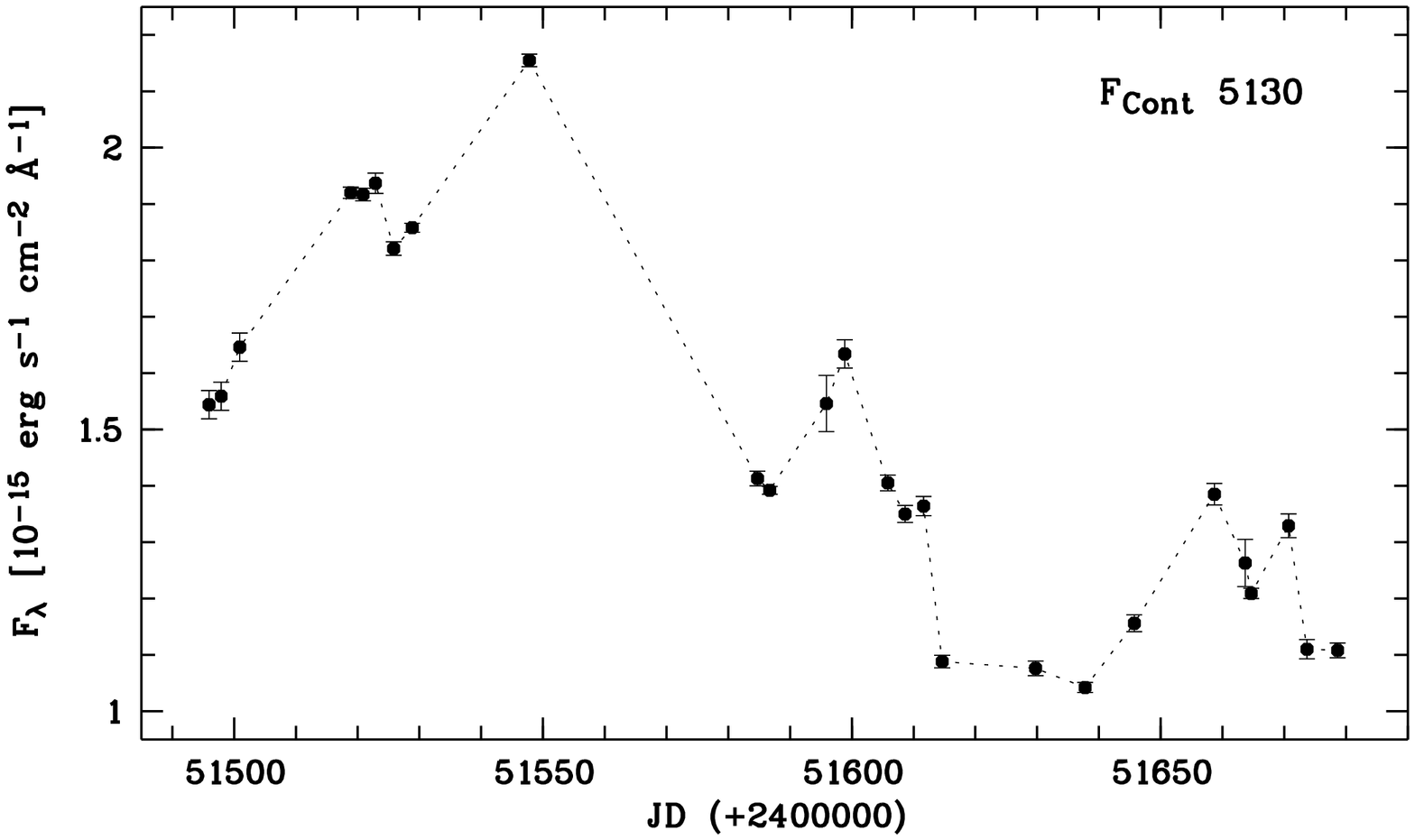}\hspace*{7mm}
       \includegraphics[bb=40 90 380 700,width=55mm,height=85mm,angle=270]
{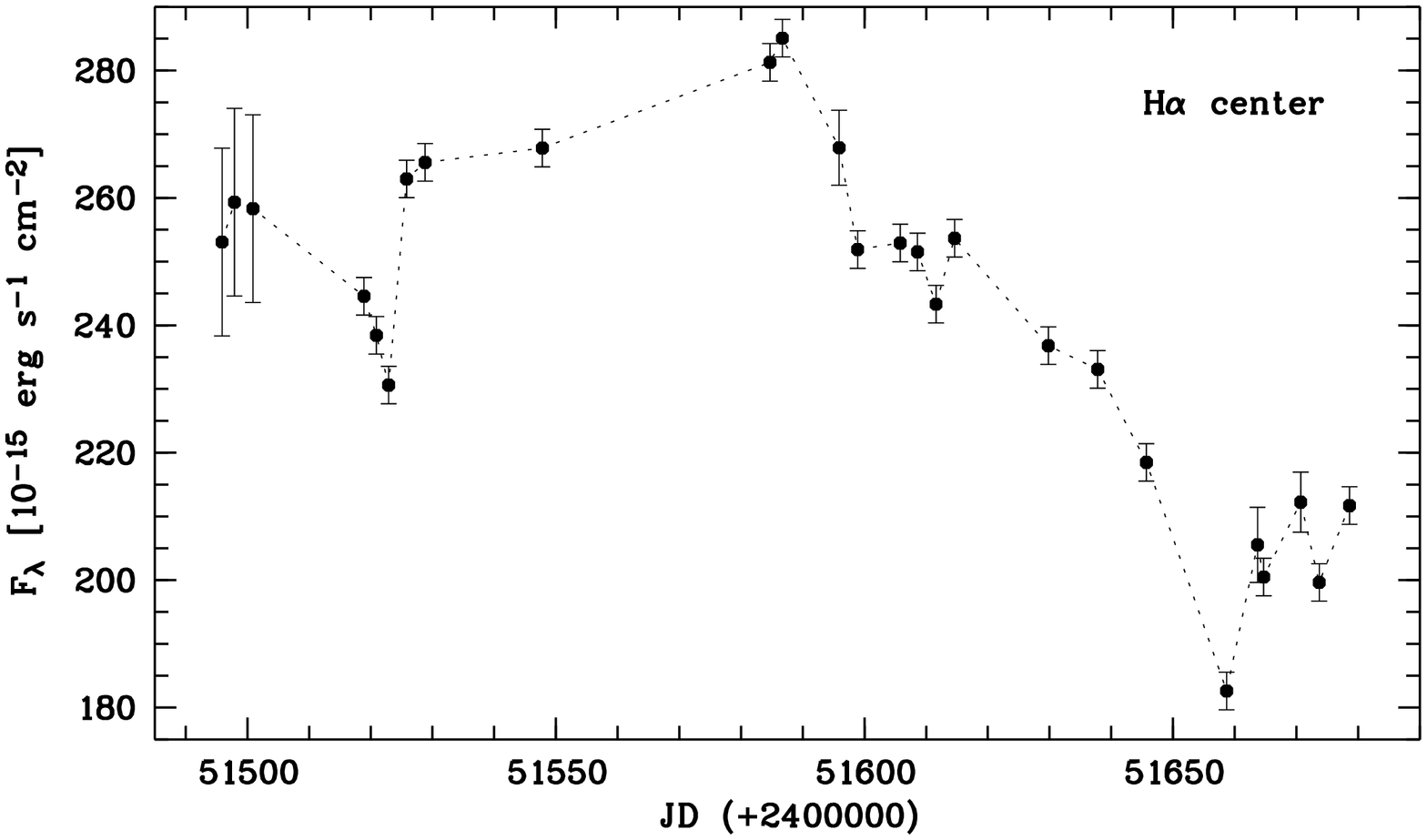}}
 \hbox{\includegraphics[bb=40 90 380 700,width=55mm,height=85mm,angle=270]
{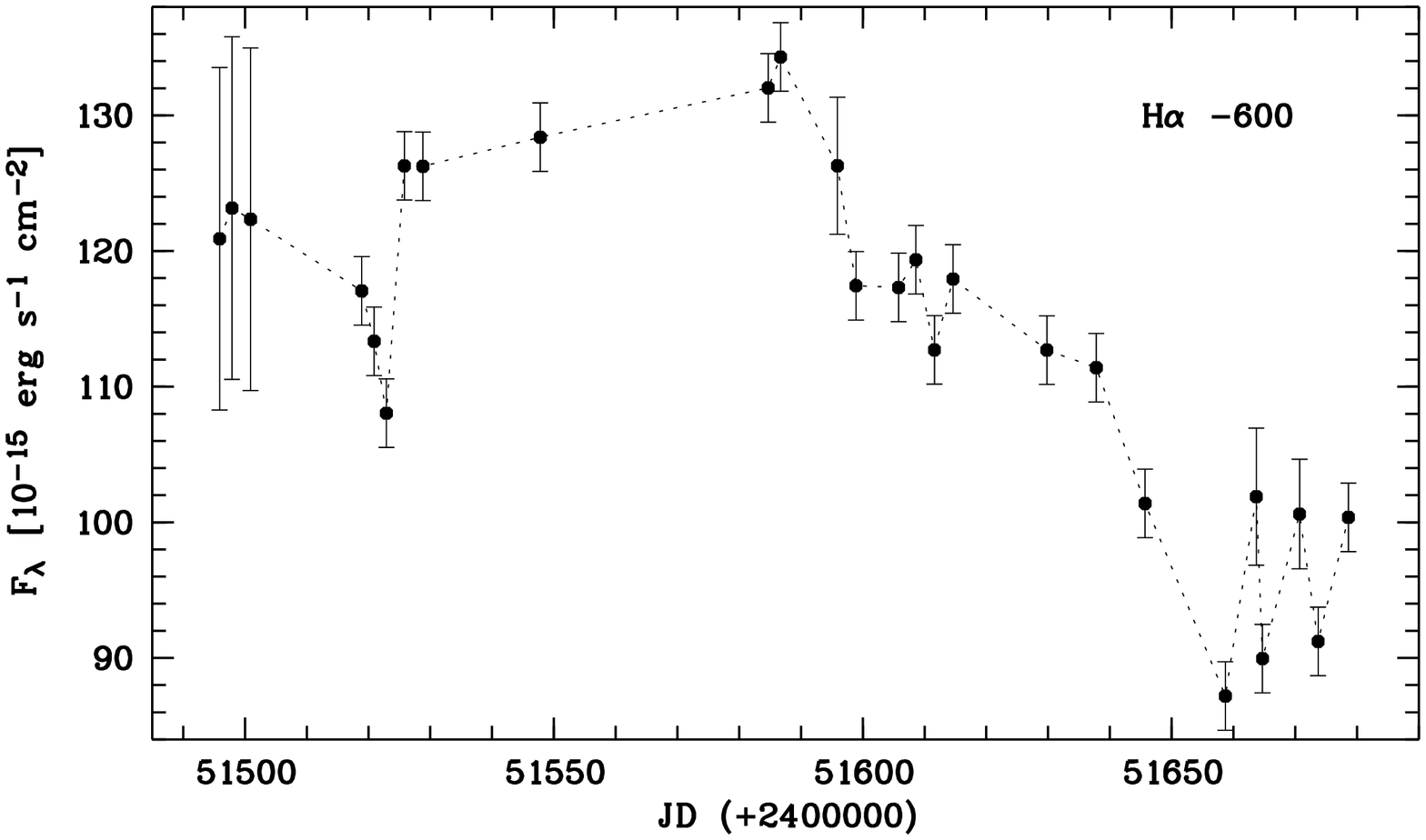}\hspace*{7mm}
       \includegraphics[bb=40 90 380 700,width=55mm,height=85mm,angle=270]
{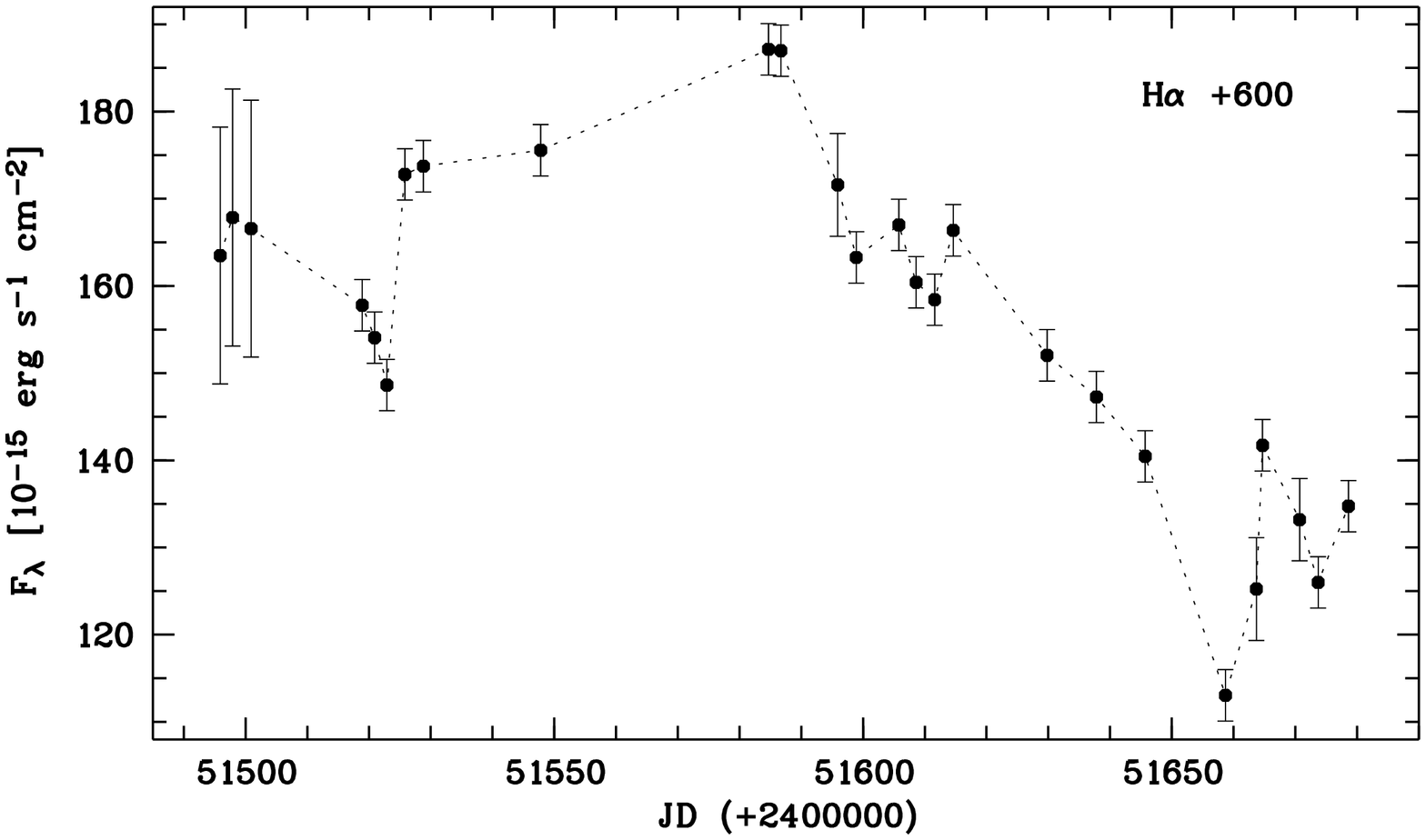}}
 \hbox{\includegraphics[bb=40 90 380 700,width=55mm,height=85mm,angle=270]
{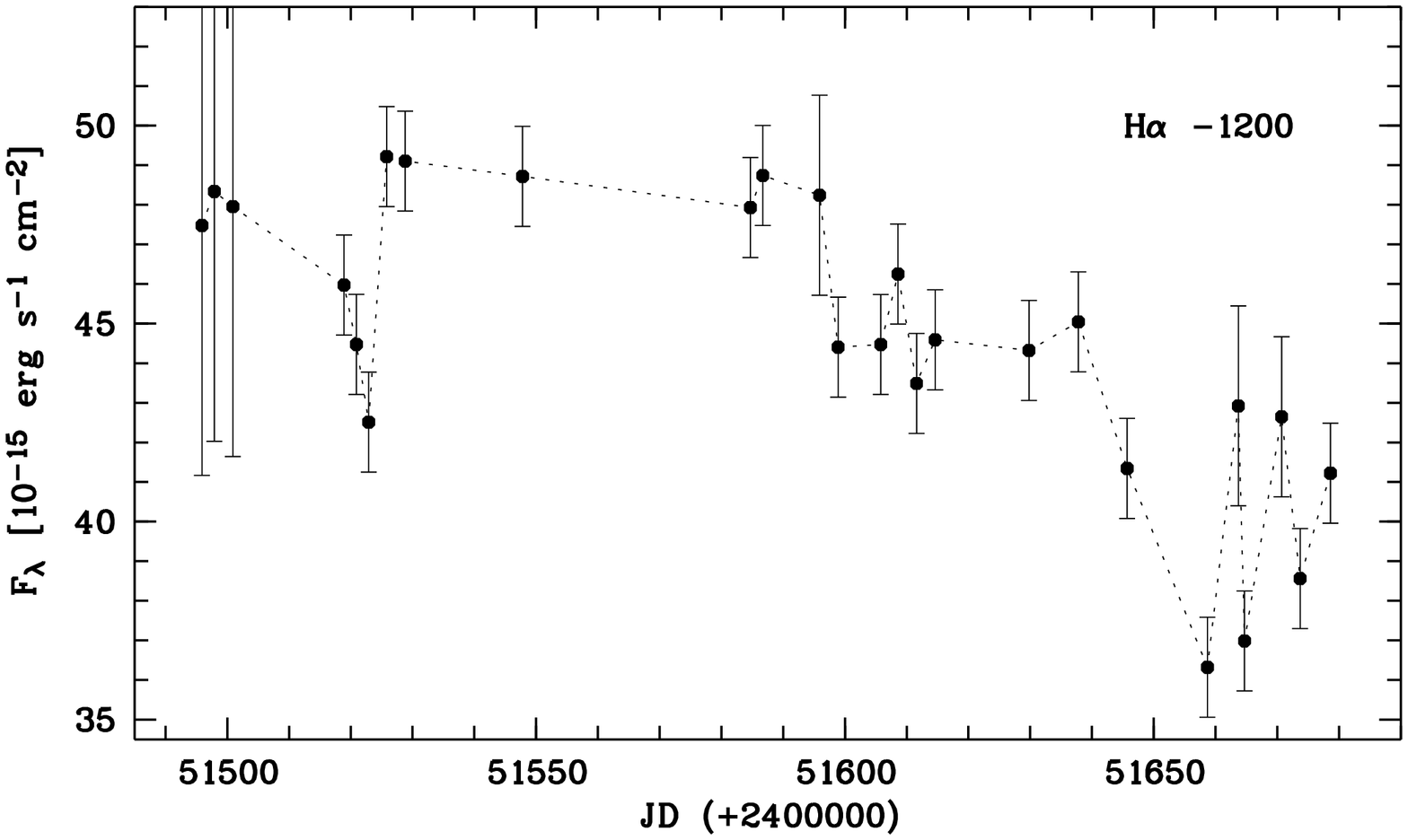}\hspace*{7mm}
       \includegraphics[bb=40 90 380 700,width=55mm,height=85mm,angle=270]
{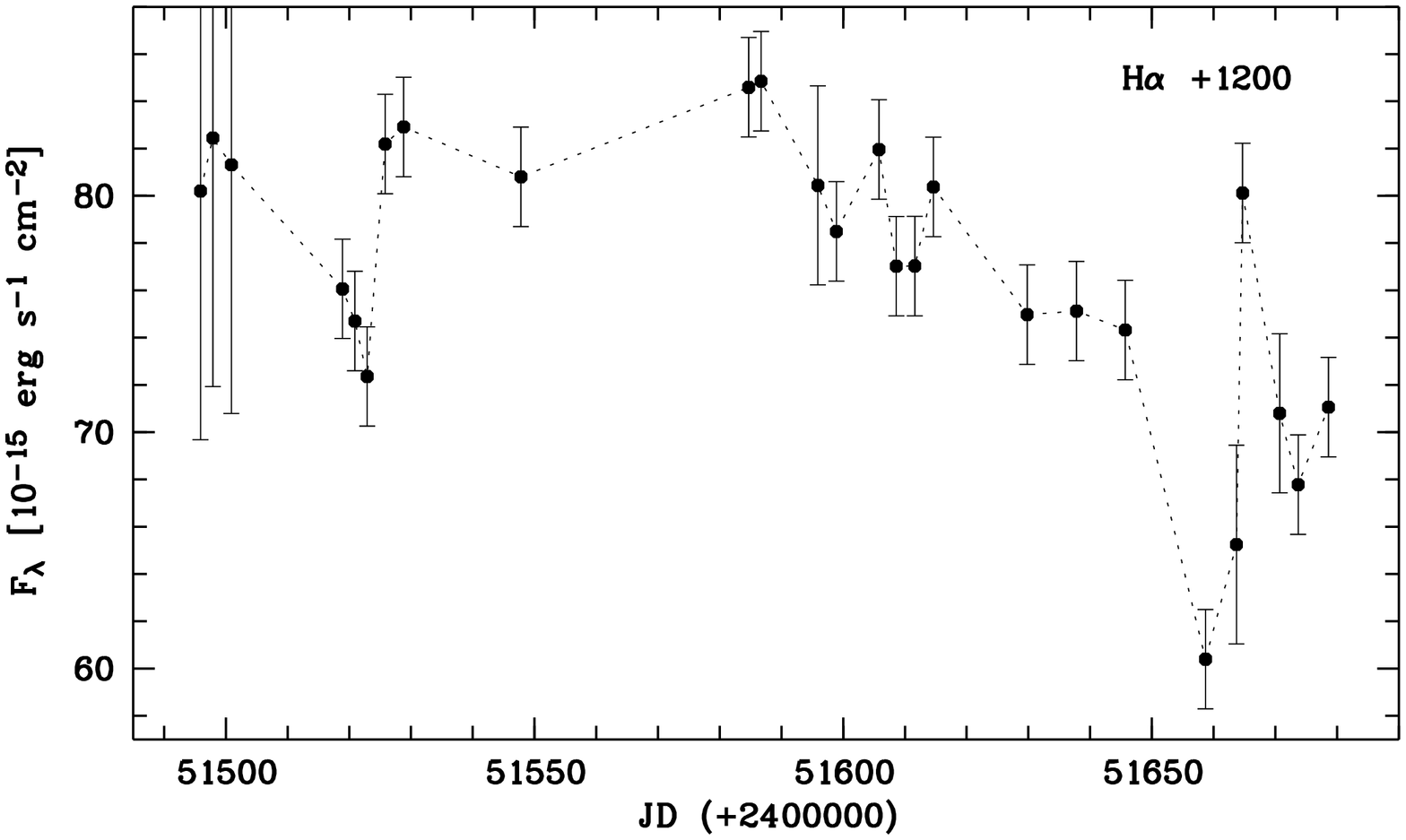}}
 \hbox{\includegraphics[bb=40 90 380 700,width=55mm,height=85mm,angle=270]
{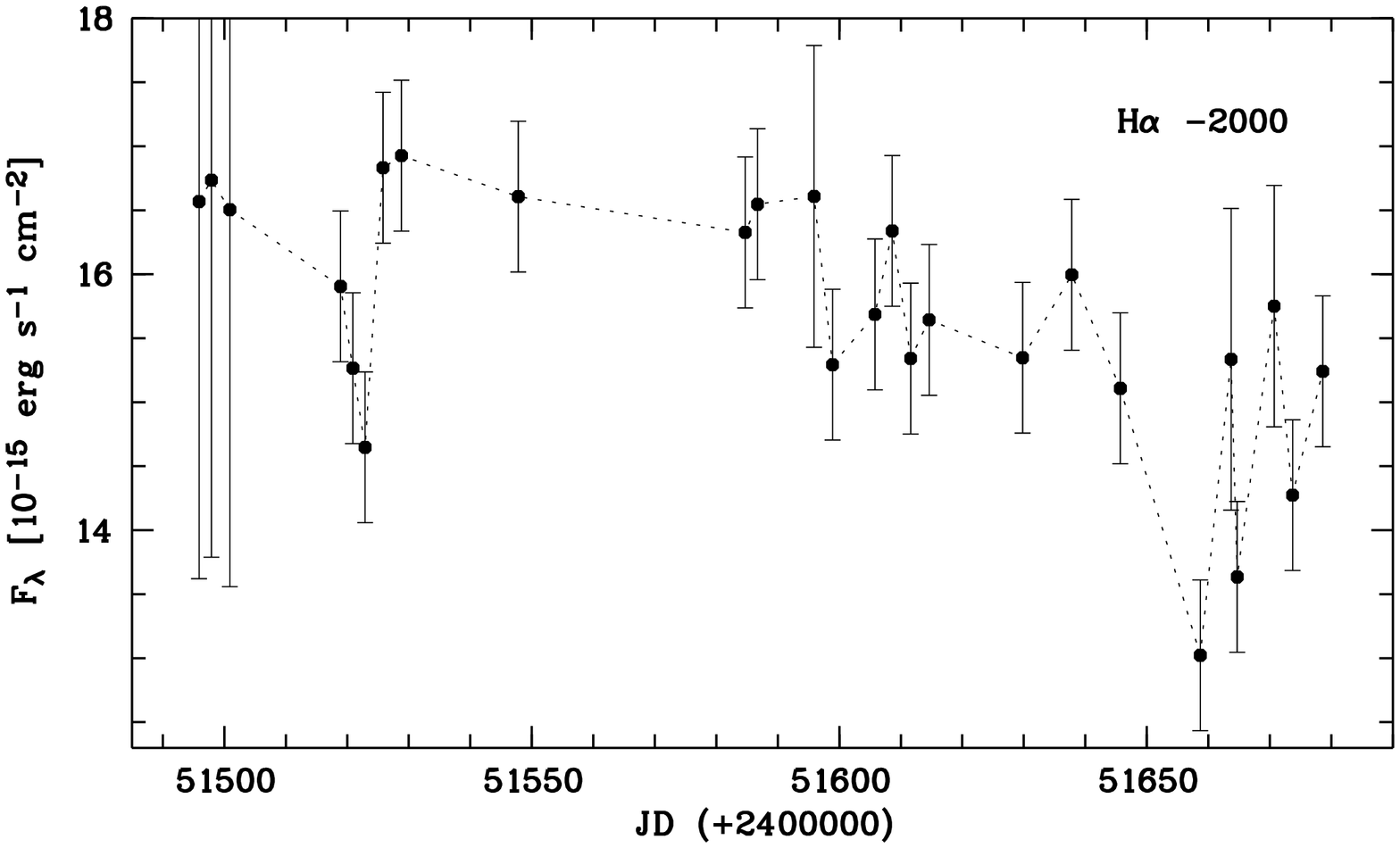}\hspace*{7mm}
       \includegraphics[bb=40 90 380 700,width=55mm,height=85mm,angle=270]
{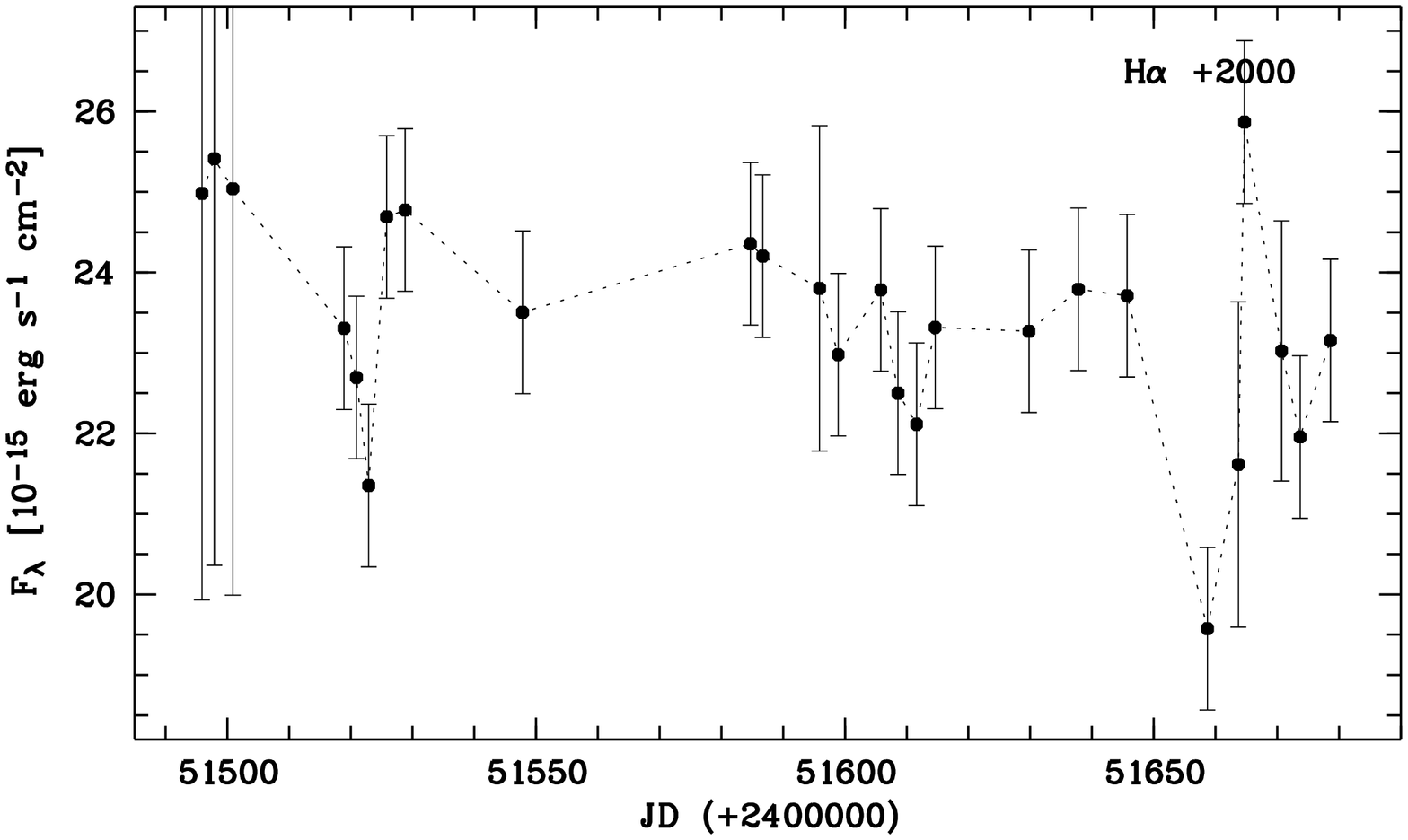}}
       \vspace*{5mm}
  \caption{Light curves of the continuum, of the H$\alpha$ line center
 as well as of different blue and red H$\alpha$ line wing segments
 ($v$ = $\pm$ 600, 1200, 2000 km/s, $\Delta$$v$ = 400 km/s) of Mrk\,110.}
% and 4265\,\AA\
%
\end{figure*}

The light curves of the segments in each line profile
are remarkably different.
There is the general trend that the pattern of the 
light curves varies
as a function of distance to line center.
On the other hand, corresponding light curves of identical 
red and blue segments are very similar. 
One can see immediately (Figs.~3 to 6)
 that the outer line wings follow closer the
continuum light curve than the inner line wings.
The errors given in the light curves indicate absolute errors.
Relative errors
between light curves of the different segments
are smaller by about 50$\%$.
\subsection{Velocity delay maps of the Balmer and Helium lines}
We computed cross-correlation functions (CCF) of all
line segment ($\Delta$$v$ = 400 km/s) light curves
with the 5100\AA\ continuum light curve.
For details of the method see Papers I, II.

The derived delays of the segments are shown in 
Fig.~7 as function of distance to the line center.
These velocity delay maps are presented in gray scale
 for the H$\beta$, HeI$\lambda$5876,
HeII$\lambda$4686,
H$\alpha$, and H$\gamma$ lines.
The solid lines show contour lines of the correlation coefficient
at levels between .800 and .925.
\begin{figure*}
 \includegraphics [width=88mm]{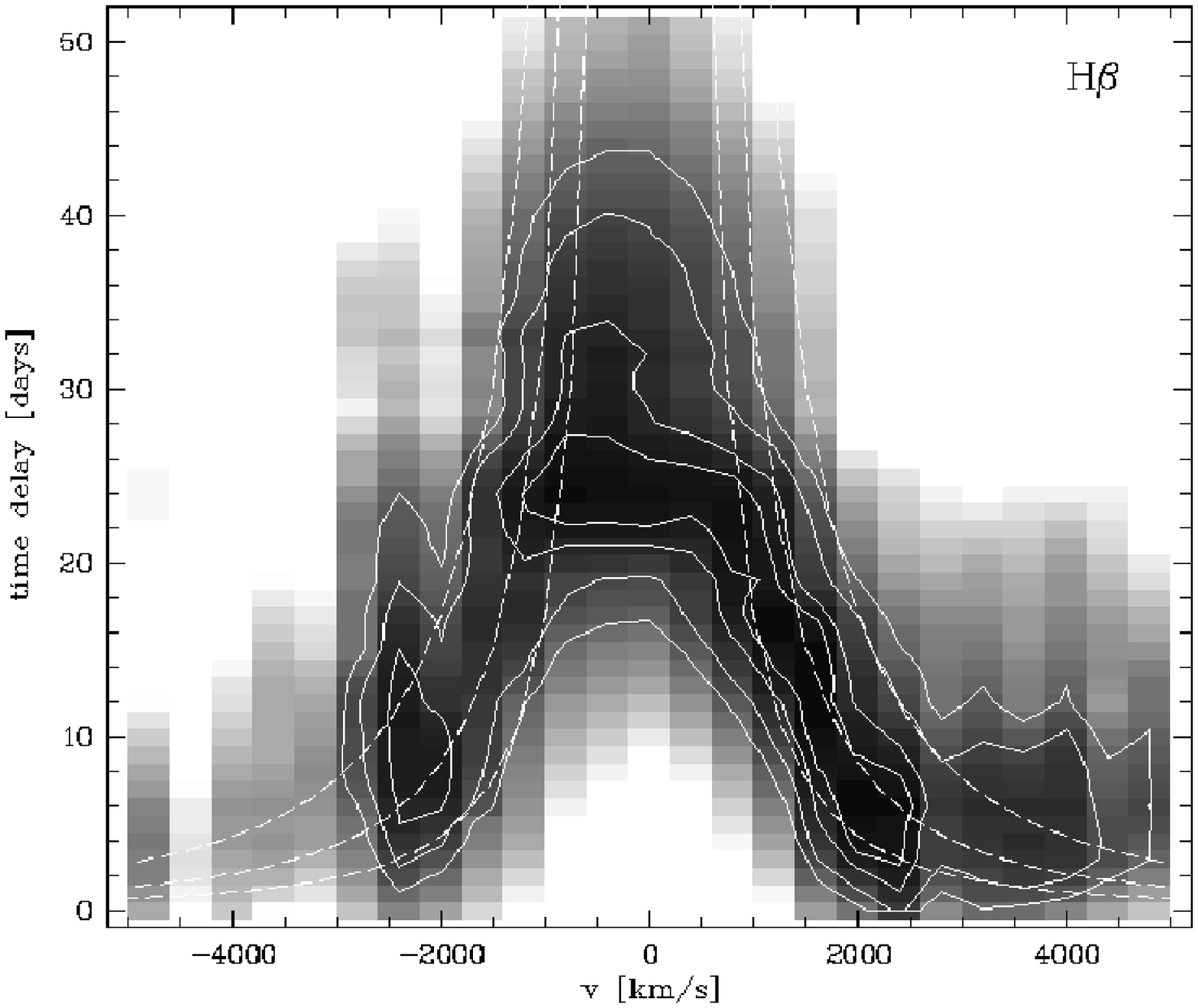}\hspace*{5mm}
 \includegraphics [width=88mm]{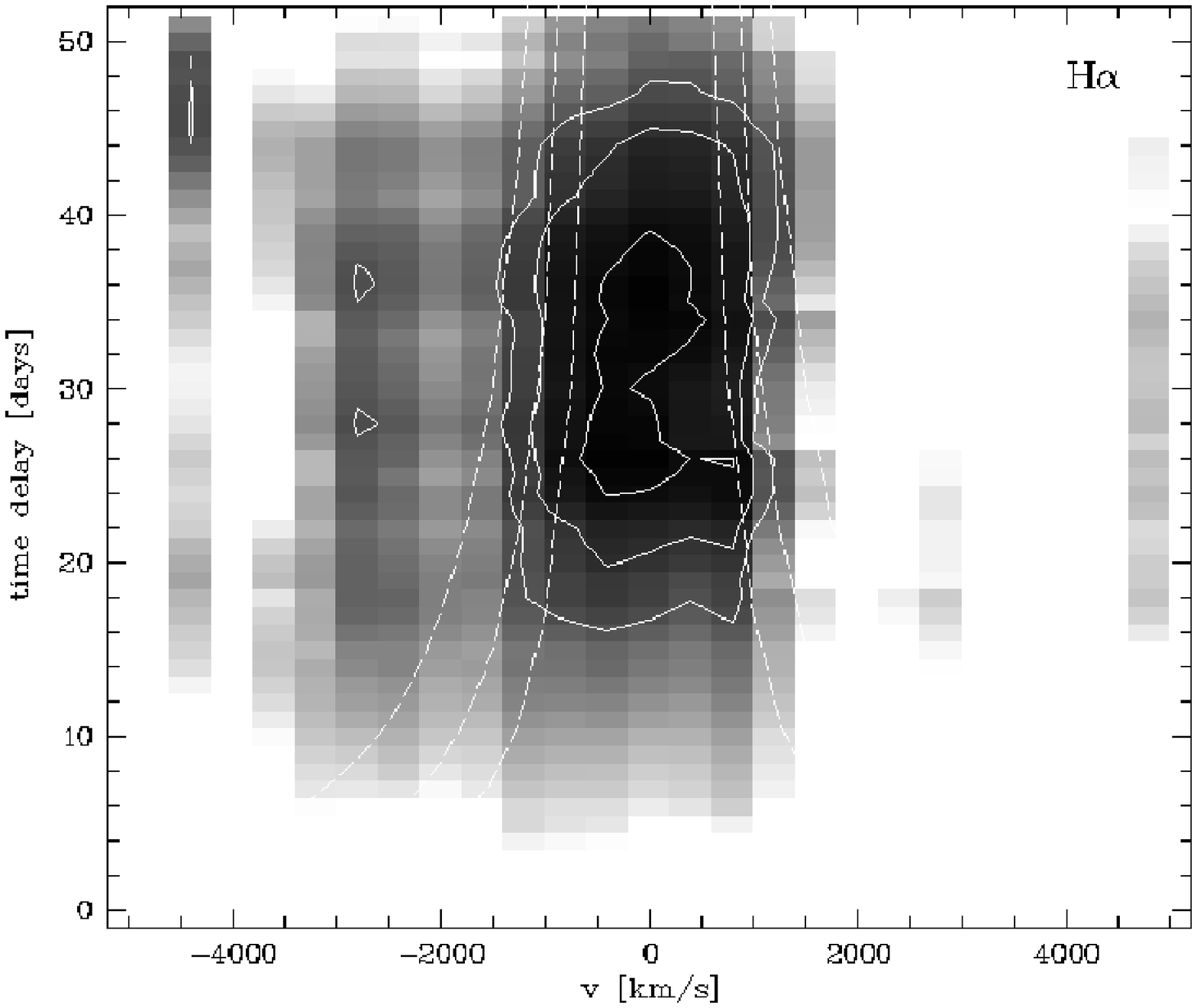}
 \includegraphics [width=88mm]{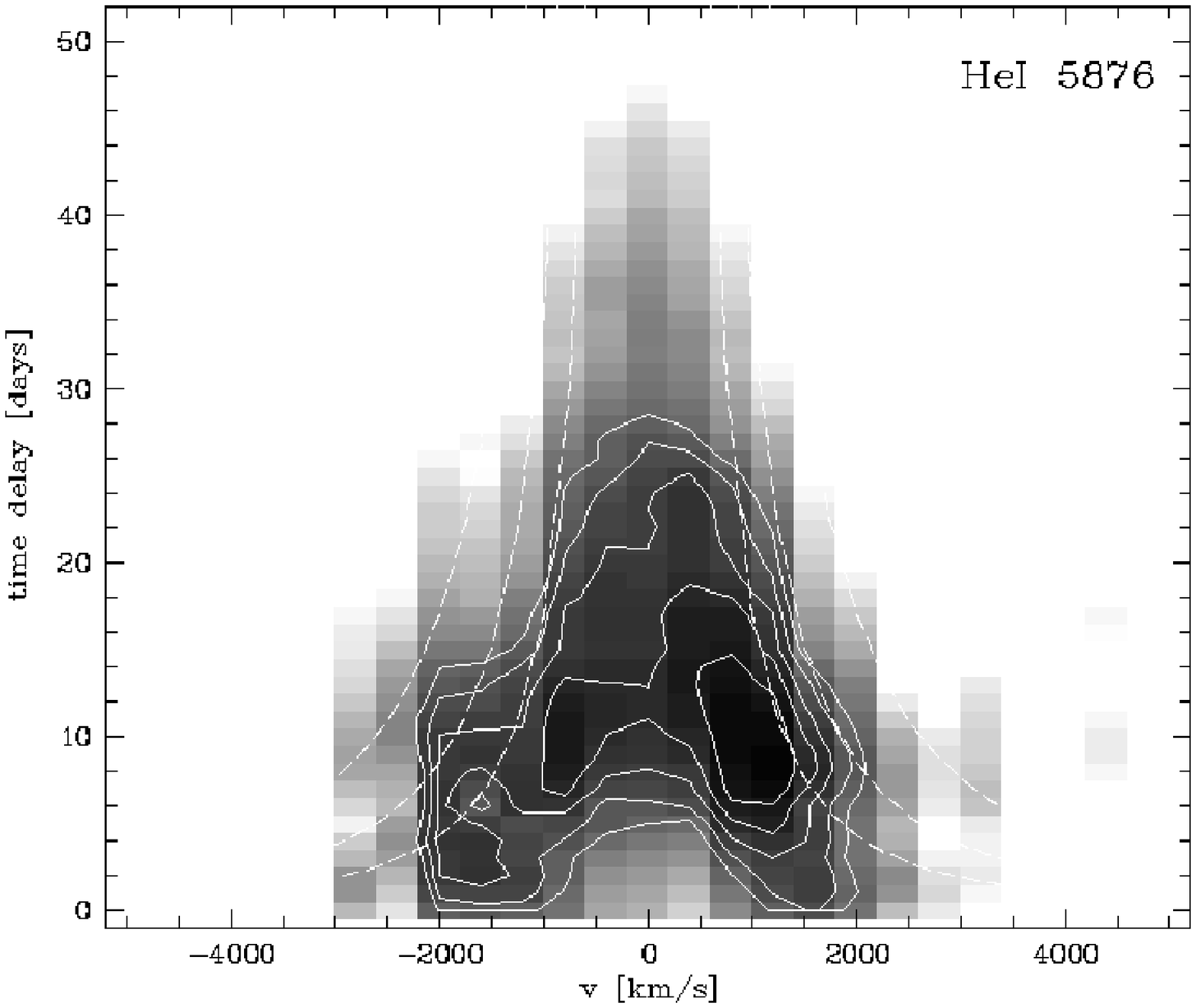}\hspace*{5mm}
 \includegraphics [width=88mm]{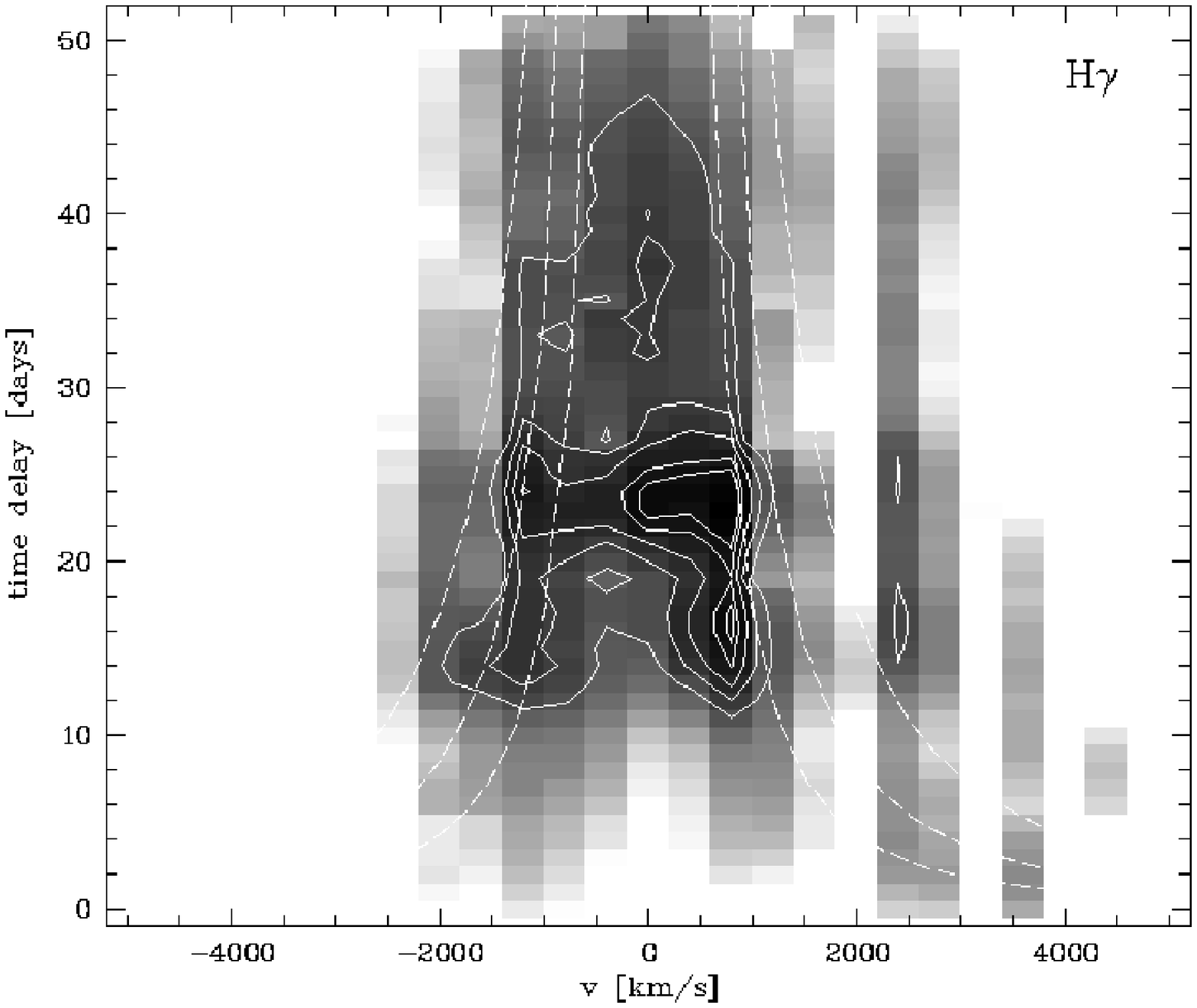}
 \includegraphics [width=88mm]{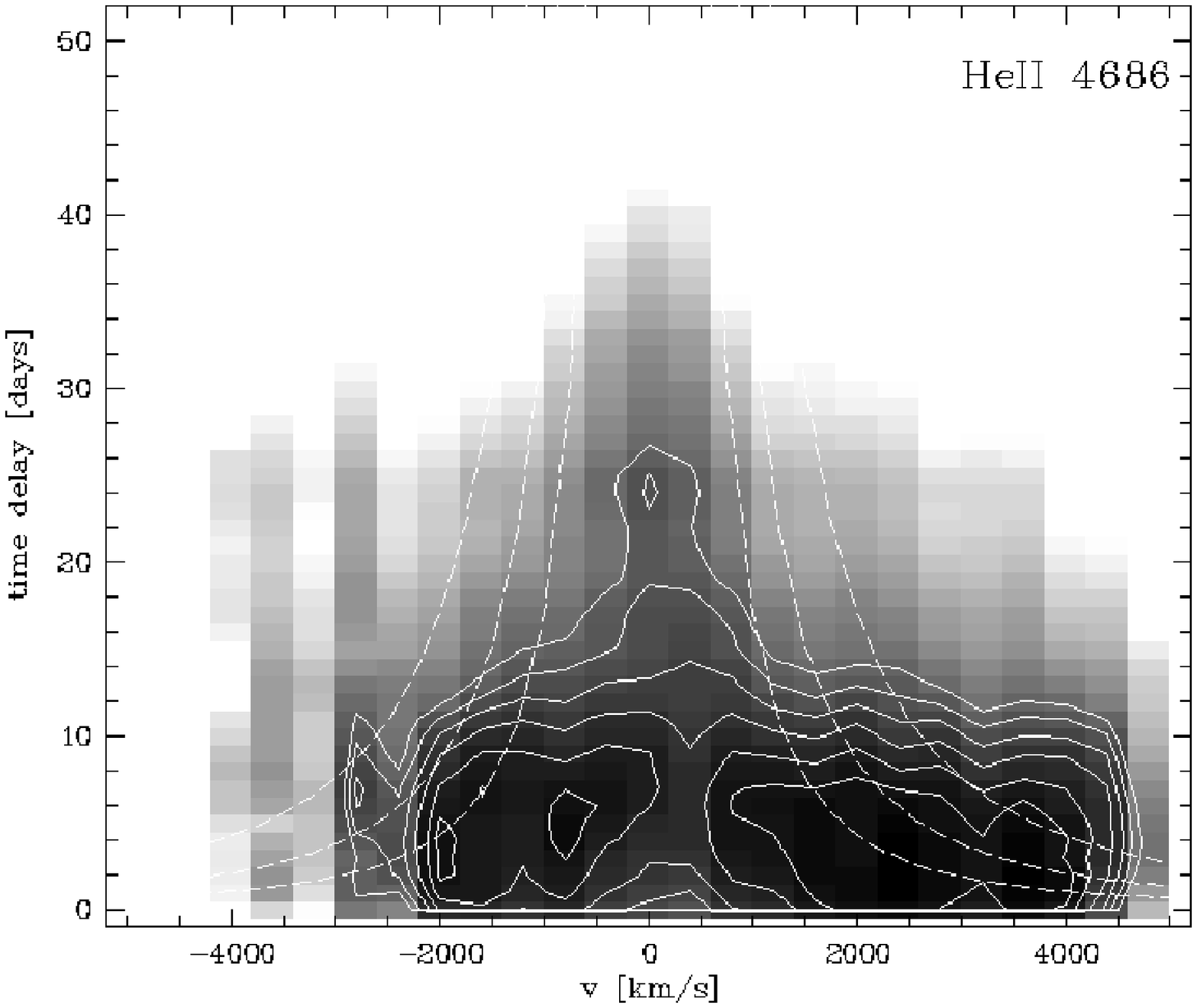}\hspace*{5mm}
  \caption{The 2-D CCFs($\tau$,$v$) show the correlation of the Balmer and 
Helium line segment light curves with the continuum light curve
as a function of velocity and time delay (grey scale).
Contours of the correlation coefficient are overplotted at levels between
.800 and .925 (solid lines).
The dashed curves show computed escape velocities for
central masses of 0.5, 1., 2. $\times\ 10^7 M_{\odot}$ (from bottom to top).
}

\end{figure*}

%
%
%

%\begin{figure*}
% \hbox{\includegraphics[bb=37 0 576 625,width=50mm]{MS3785f7a.ps}\hspace*{7mm}
%       \includegraphics[bb=37 0 576 625,width=88mm]{MS3785f7b.ps}}\vspace*{-15mm}
% \hbox{\includegraphics[bb=37 0 576 625,width=88mm]{MS3785f7c.ps}\hspace*{7mm}
%       \includegraphics[bb=37 0 576 625,width=88mm]{MS3785f7d.ps}}\vspace*{-15mm}
% \hbox{\includegraphics[bb=37 0 576 625,width=88mm]{MS3785f7e.ps}\hspace*{7mm}}

% \hbox{\includegraphics[bb=37 167 576 625,width=88mm]{MS3785f7e.ps}\hspace*{7mm}}
%       \vspace*{5mm}
%  \caption{The 2-D CCFs($\tau$,$v$) show the correlation of the Balmer and 
%Helium line segment light curves with the continuum light curve
%as a function of velocity and time delay (grey scale).
%Contours of the correlation coefficient are overplotted at levels between
%.800 and .925 (solid lines).
%The dashed curves show computed escape velocities for
%central masses of 0.5, 1., 2. $\times\ 10^7 M_{\odot}$ (from bottom to top).
%}
%
%H$\beta$ Contours .85, .88, .91, .925 %gray: 0.70-0.95
%HeII$\lambda$4686 .83, .85, .87, .89, .91, .927 %gray: 0.65-0.95
%HeI$\lambda$5876 .85, .87, .89, .907, .925 %gray: 0.70-0.9
%H$\gamma$ .82, .85, .87, .89, .90 %gray: 0.62-0.92
%H$\alpha$ .80, .85, .903 %gray: 0.55-0.92
%

%\end{figure*}
%
%
The H$\gamma$ line
and the redshifted H$\alpha$ line are heavily contaminated
by other emission and/or absorption lines
as had been said before.
Therefore, the correlation coefficient is smaller in the wings of these
lines.
From now on we will consider only the H$\beta$,
HeI$\lambda$5876, and  HeII$\lambda$4686 lines
for a more detailed discussion.

The light curves of the line center
are mostly delayed with respect to continuum variations.
The outer line wings respond
much faster to continuum variations respectively
than the inner line profile segments.
Comparing the 2-D CCFs of the different lines with each other
one can identify 
a clear stratification within the broad line region.
The same trend has been seen before in
the integrated lines.
The HeII$\lambda$4686 line responds first to continuum variations followed 
by HeI$\lambda$5876 and finally H$\beta$.
The correlation coefficients in the line wings as well as in the line
centers of the H$\beta$ and He lines are very similar
 although the intensities in the line wings
are obviously smaller.
But one has to keep in mind that
 the line wings originate closer to the nuclear 
ionizing source than the line centers.

The 2-D CCF($\tau$,$v$) is
mathematically very similar to a 2-D response function $\Psi$ (Welsh
 \cite{welsh01}).
In the next section the observed velocity delay maps
are compared in more detail
with model calculations of echo images from the BLR.
The BLR
Keplerian disk model of 
Welsh \& Horne (\cite{welsh91}, Fig.~1c)
shows remarkable coincidences with our observations.
The three dashed lines overplotted additionally
in Fig.~7 represent escape velocities computed for
central masses of 0.5, 1., 2. $\times\ 10^7 M_{\odot}$ (from bottom to top).\\
\subsection{Delay between blue and red line wings}
As one part of our analysis
we calculated 
cross-correlation functions (CCF) of the
blue line wings segments with respect to the red ones
to determine more accurately their relative response.
This was done
for all the strong emission lines.
We computed the relative delays in the line wings 
in velocity segments of $\Delta$$v$ = 400 km/s width respectively
beginning at $\pm$ 200 km/s. 
Figs.~8 to 10 show the delay of the blue line wings with respect to
the red ones for the H$\beta$, HeI$\lambda$5876 and HeII$\lambda$4686
lines (in gray scale).
Contours of the correlation coefficient are overplotted at levels between
.85 and .97 (solid lines).
The center of the CCFs is indicated by the short dashed lines.
The long dashed lines
show the centroid of the uppermost 10\% of the CCF
\begin{figure}
\includegraphics[width=88mm]{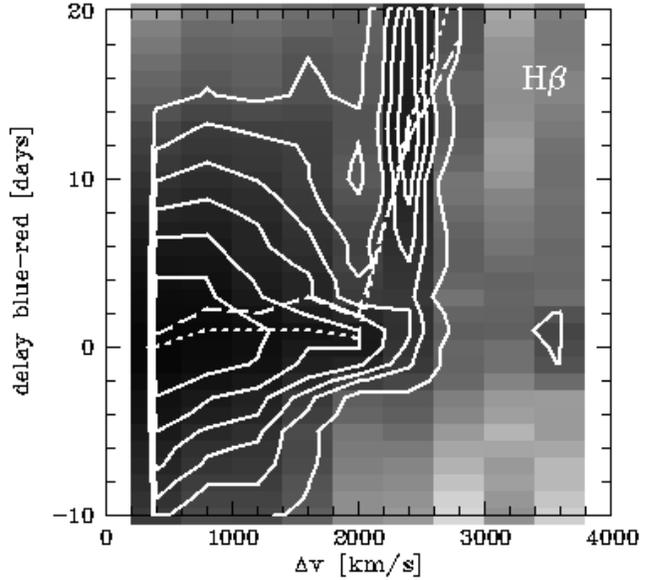}
\caption{Time delay $\tau$ of the blue H$\beta$
line wing with respect to the
red one as a function of distance to the line center.
Contours of the correlation coefficient are overplotted at levels between
.85 and .97 (solid lines).
The short dashed line gives the center of the CCF. The long dashed line
gives the centroid of the uppermost 10\% of the CCF.}
%
%H$\beta$ %gray: 0.55-0.99
%
\end{figure}
\begin{figure}
\includegraphics[width=88mm]{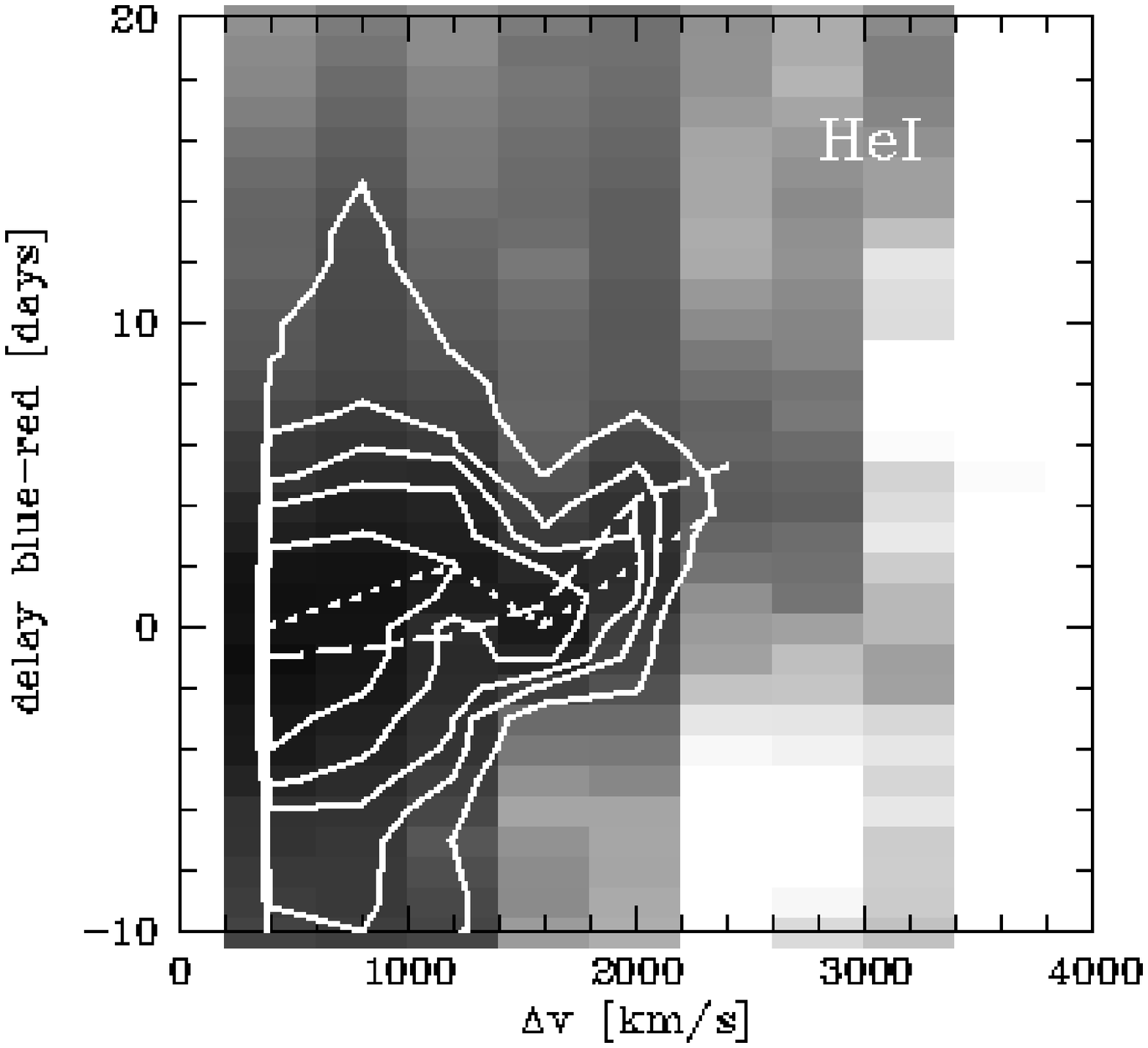}
\caption{Time delay $\tau$ of the blue HeI$\lambda$5876
line wing with respect to the
red one as a function of distance to the line center.
Contours of the correlation coefficient are overplotted at levels between
.85 and .97 (solid lines).
The short dashed line gives the center of the CCF. The long dashed line
gives the centroid of the uppermost 10\% of the CCF.}
%
%HeI$\lambda$5876 %gray: 0.55-0.99
%
\end{figure}
\begin{figure}
\includegraphics[width=88mm]{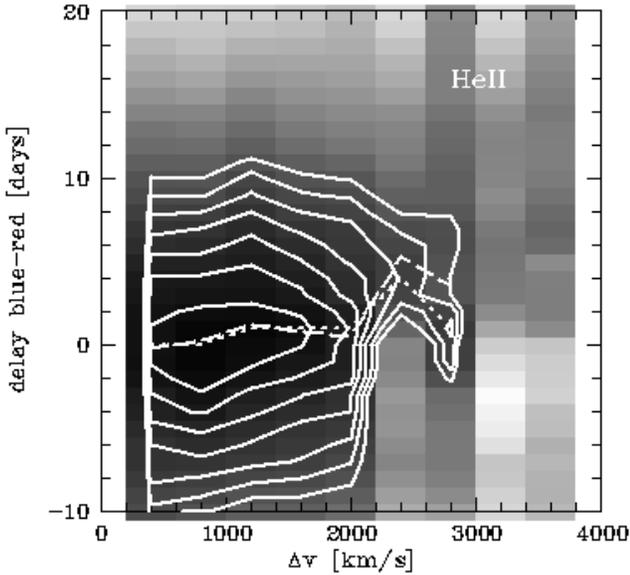}
\caption{Time delay $\tau$ of the blue HeII$\lambda$4686
line wing with respect to the
red one as a function of distance to the line center.
Contours of the correlation coefficient are overplotted at levels between
.85 and .97 (solid lines).
The short dashed line gives the center of the CCF. The long dashed line
gives the centroid of the uppermost 10\% of the CCF.}
%
%HeII$\lambda$4686 %gray: 0.55-0.99
%
\end{figure}

The same trend is to be seen in all emission lines:
The blue line wings show a delayed response
with respect to the red ones
at intermediate distances from the line center
% wing segments at distances of
($\Delta$v~=~500 -- 2000 km/s)
by one to two days. 
Furthermore, there is a trend
that the relative delay of the blue wings
increases with distance
to the line center.
\section{Discussion}
\subsection{BLR stratification derived from integrated lines}

The integrated emission lines in Mrk\,110 respond
with different delays to continuum variations 
 as a function
of ionization degree. This  indicates a 
stratification in the broad-line region (BLR) of Mrk\,110
 (Kollatschny et al., 2001).
Fig.~2 shows a clear correlation between  
time lag $\tau$ of different integrated emission lines on the one hand
 and their corresponding line widths $v$ on the other hand.
The correlation is of the form
\[ \tau \propto v_{FWHM}^{-2} \] 
for distances $R~=~c\tau$ from the central ionizing source.
 All emission 
line data are consistent with a virial mass of 
\[ M= 1.8\pm0.3\cdot10^{7} M_{\odot} . \]
The higher ionized broader emission lines originate closer to the 
central ionizing source than the lower ionized lines.

\subsection{BLR geometry and kinematics}

We sliced the observed emission line profiles into velocity bins of
$\Delta$$v$ = 400 km/s
and correlated all the individual light curves with the continuum light 
curve.
First results of this investigation
 have been published for the H$\beta$ line indicating that
different line segments originate at different distances from
the ionizing center.

The velocity resolved profile variations of all investigated broad lines
(Fig.~7) show the
same characteristics. The
line segments in each profile originate at different distances from the
central ionizing source.
The observed trend in Mrk\,110 is that the outer emission
line wings originate at small radii
from the central supermassive black hole.
%to the galactic center than the innermore line regions 
%is an indication for the dominance of
This suggests strong
rotational and or turbulent 
motions in the BLR (see however section 4.3).

Comparing in more detail
the observed velocity-delay pattern with BLR model
calculations 
(Welsh \& Horne \cite{welsh91}; Perez et al. \cite{perez92};
 O'Brien et al. \cite{obrien94}) we can rule out
that radial inflow or outflow motions -- including biconical outflow --
are dominant in the BLR of Mrk\,110.
The line wings show the shortest delay with respect to the continuum
and react nearly simultaneously.
Furthermore,
no short delays of the central region in the Balmer
and He lines are observed.
This is expected in
 spherical BLR models
with chaotic virial velocity field or randomly oriented Keplerian orbits.
Therefore we can rule out a dominance of this
kind of velocity field in Mrk\,110.
On the other hand Keplerian disk BLR models reproduce exactly 
the observed velocity-delay pattern i.e.
the faster response of both line wings compared to the line center.

A direct comparison of the two-dimensional echo maps of the
H$\beta$, HeI$\lambda$5876, and HeII$\lambda$4686 lines 
with a theoretical echo image of an Keplerian disk 
(Welsh \& Horne \cite{welsh91}, Fig.~1c) is
intriguing. Our observed 2D pattern of the different lines in Mrk\,110
has been predicted in theoretical models which calculated the
contibution from line emitting material at different
radii in the BLR. The HeII$\lambda$4686 line originates at
 radial distances of 3 -- 4 light days only
while H$\beta$ originates at distances of about 30 light days.

A coarse estimate of the inclination angle of the accretion disk
can be made by comparing the echo maps (Fig.~7) with disk models viewed  
at different inclination angles (Welsh \& Horne \cite{welsh91}, Fig.~5).
We observe no short delays at the inner line regions.
This excludes inclination angles 
of the accretion disk larger than 50${^\circ}$ in Mrk\,110.
At the limit of a face-on disk one expects even narrower line profiles 
than those we observe in Mrk\,110.
Therefore, a best estimate of the accretion disk inclination angle in Mrk\,110
is: 30${^\circ}$$\pm$20${^\circ}$.

Other authors published
further indications 
for a disk-like configuration of the broad-line region in AGN
based on theoretical models or observational data
(Bottorff et al. \cite{bottorff97},  
Collin-Souffrin et al. \cite{collin88},
Elvis \cite{elvis00},
K\"{o}nigl \& Kartje \cite{koenigl94}).

Velocity-delay maps have been published for the Balmer
lines in NGC\,5548 (Kollatschny \& Dietrich \cite{kollatschny96}) and
NGC\,4593 (Kollatschny \& Dietrich \cite{kollatschny97}) before. 
Although the quality of their spectral data was not as good as that of
this campaign
they could demonstrate the
same basic trend as that seen in Mrk\,110:
the outer line wings respond faster
to continuum variations than
the line center.
Furthermore, the red wings responded slightly faster and
stronger than the blue ones in theses galaxies, too -- see next section. 
In the UV wavelength range the variability behaviour of the prominent
 CIV$\lambda$1550 line has been investigated
in a few galaxies.
 Again 
there are hints for a stronger and faster response of the red wing in 
NGC\,5548 (Chiang \& Murray \cite{chiang96}, Bottorff et al. \cite{bottorff97})
 and
NGC\,4151 (Ulrich \& Horne \cite{ulrich96})
in comparison to the blue one. Unfortunately,
the CIV$\lambda$1550 emission lines
are strongly affected by a central absorption line blend
in both galaxies.

\subsection{Accretion disk wind}

In this section I will put again the main emphasis on the
velocity-delay data of the H$\beta$,
HeII$\lambda$4686, and HeI$\lambda$5876 lines (Figs.~7 to 10).
A careful inspection shows
that a second trend in these 2D velocity-delay maps is
superimposed on the primary trend
that both outer line wings respond faster than 
central line region:
The response of the red line wings is
stronger than that of the blue ones.
% in these velocity delay maps.
For all line profiles
the correlation coefficients 
of the red wing light curves with the continuum light curve 
are systematically
higher by 5\% (Fig.~7) than those ones of the blue wing. 
Furthermore, the red wings respond faster than the blue ones.
The integrated blue wings
(Figs.~7 -- 10) lag the red wings by $2^{+2}_{-1}$ days. 

An earlier response of the red line wing compared to the blue line wing
is predicted in  spherical wind and
 disk-wind models of the BLR 
(K\"{o}nigl \& Kartje \cite{koenigl94},
Chiang \& Murray \cite{chiang96},
Blandford \& Begelman \cite{blandford99}).
In these models the line emitting gas shows a radial outward
velocity component in addition to the rotation.
Also a stronger response of the red wing is expected in the models
and observed in NGC\,5548 (Chiang \& Murray \cite{chiang96}). But we could
not verify a secondary peak in our data as predicted in the model of 
(Chiang \& Murray \cite{chiang96}).
The disk outflow/wind models
are distinguished from spherical wind models that their velocity
decreases with radius (this means from the outer line wings to the
line center) rather than the other way around.
The observed delays of the blue line wings with respect to the red ones
(Figs. 8 -- 10) point at an 
accretion disk-wind  in Mrk\,110.
In particular one can see an increase of the blue-red delay towards 
the outer line wings which means that the wind velocity increases towards
the center.

Furthermore, Murray \& Chiang (\cite{murray97}) demonstrated that a
Keplerian disk wind reproduces single-peaked broad emission lines
as we see in the spectra of Mrk\,110.
\subsection{Central black hole mass}

From the integrated line intensity variations of four different
emission lines a virial mass of
\[ M= 1.8\pm0.3\cdot10^{7} M_{\odot}  \]
has been determined (Sect.3). 
We made the assumption
that the characteristic velocities of the particular emission line
regions are given by the FWHM of their rms profiles and the characteristic 
distances R are given by the centroids of the corresponding cross-correlation
 functions:
\[ M = \frac{3}{2} v^{2} G^{-1} R \] 
(e.g. Koratkar \& Gaskell \cite{koratkar91}, Kollatschny \& Dietrich
  \cite{kollatschny97}).

Additionally, we may compare the velocity-delay maps (Fig.~7) 
with escape velocity envelopes
\[ v = \sqrt{2 G M / c \tau} \] 
for different central masses.
With this method we derive a central black hole
mass of
\[ M= 1.5\pm1\cdot 10^{7} M_{\odot}  \]
in Mrk\,110.
This mass value confirms perfectly our earlier derived value.

But one has to keep in mind that
there are further systematic uncertainties in the mass determination
(e.g. Krolik \cite{krolik01}).
Due to the unknown inclination
angle of the accretion disk the derived mass may be a lower limit only. 
Our observed velocity-delay maps do not drop down near the line 
center (Fig.\ 3) as is expected from model 
calculations for edge-on disk models (Welsh \& Horne \cite{welsh91}; 
O'Brien et al. \cite{obrien94}).
This is a hint for a small inclination angle of the
accretion disk in Mrk\,110. 

Ferrarese et al. \cite{ferrarese01} measured the stellar velocity dispersion
in the CaII triplet lines of the host galaxy in Mrk\,110.
 Their derived velocity
dispersion corresponds to a central mass of only
\[ M = 3\cdot 10^{6} M_{\odot} \]
in their black hole mass vs. velocity dispersion 
diagram. This value
 is a factor of 5 below our black hole mass value we derived with 
other methods for this galaxy. But one has to consider that
Mrk\,110 is a galaxy system in the late stage of merging
with respect to its morphology. 
 Therefore, 
the derived central stellar velocity dispersion in the host galaxy of Mrk\,110
might by contaminated heavily  by this interaction/merging effect.
\subsection{Schematic BLR model}
We are now in the position of
generating a schematic model of the innermost AGN region in Mrk\,110
from all our data. The result is
shown in Fig.~11.
\begin{figure}
\includegraphics[width=88mm]{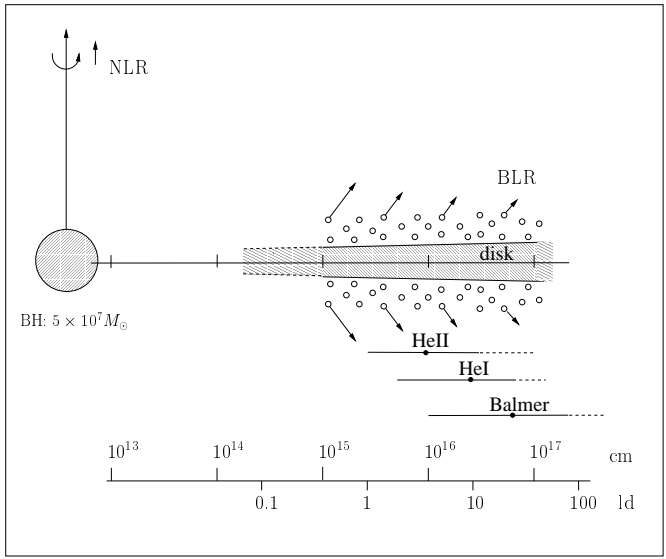}
\caption{Schematic BLR model of Mrk~110.}
\end{figure}
We could demonstrate that the broad emission lines originate in the
wind of an accretion disk.
The distances of the line emitting regions from
the central ionizing source are shown on a logarithmic scale.
A central black hole mass of
$M= 1.8\pm1\cdot 10^{7} M_{\odot}$
is a lower limit only
because of the poorly known inclination angle of the accretion disk. This
value is of the order of 30${^\circ}$.
Therefore, we draw a Schwarzschild radius for a black hole mass of
$M= 5\cdot 10^{7} M_{\odot}$ in Fig.~11. 
\section{Summary}
We analyzed carefully Balmer and Helium 
emission line profile variations which we derived from our monitoring campaign 
on the Seyfert 1 galaxy Mrk\,110.
Three clear trends are to be seen
in our generated velocity-delay maps:\\
- The outer red and blue line wings of the emission profiles
respond nearly simultaneously 
with respect to variations of the ionizing continuum
and much
faster than the inner regions in the emission lines.
A detailed comparison of 
our observed velocity-delay maps with theoretical models of other authors
points clearly to 
an accretion disk where the broad emission lines originate.
\\
-There is the general trend starting from the Balmer lines upon the
HeI line up to the HeII lines that 
the higher ionized lines
respond systematically faster.
This is to be seen in
the integrated lines on the one hand
as well as in the velocity resolved line profiles on the other hand.
It indicates an ionization stratification in the broad-line
region.
\\
- The response of the red line wing with respect to the ionizing
continuum variations
 is systematically stronger and slightly
faster compared to the blue line wing. This finding
is the signature of an
accretion disk wind in the broad-line region.
\\
We derived from the integrated line intensity variations
as well as
from the line profile variations a central
black hole mass of  M = $1.8\cdot10^{7} M_{\odot}$.
This value is a lower
limit because of the unknown accretion disk orientation.
% not considered projection effect of
%the accretion disk.
%
%
%
\begin{acknowledgements}
      WK thanks the UT Astronomy Department for warm hospitality during
      his visit. He thanks K. Bischoff, M. Bottorff, and M. Zetzl
 for valuable comments.
      Part of this work has been supported by the
      \emph{Deut\-sche For\-schungs\-ge\-mein\-schaft, DFG\/} grant
      KO 857/24 and DAAD.
\end{acknowledgements}


\begin{thebibliography}{}

  \bibitem[1999]{blandford99} Blandford, R.D.,\& Begelman, M.C. 1999,
       MNRAS, 303, L1

  \bibitem[1982]{blandford82} Blandford, R.D.,\&  Payne, D.G. 1982,
       MNRAS, 199, 883

  \bibitem[1997]{bottorff97} Bottorff, M., Korista, K.T., Shlosman, I. \&
        Blandford, R.D.
       1997, ApJ, 479, 200

  \bibitem[1996]{chiang96} Chiang, J., \& Murray, N. 1996, ApJ, 466, 704

  \bibitem[2002]{cohen02} Cohen, M.H. \& Martel A.R. 2002, in:
       Crenshaw D.M. et al. (eds.): 
       Mass outflow in AGN, ASP Conf.Ser. 255, p.255

  \bibitem[1988]{collin88} Collin-Souffrin, S., Dyson, J.E., McDowell, J.C.,
      \& Perry J.J. 1988, MNRAS, 232, 539

  \bibitem[2000]{elvis00} Elvis, M. 2000, ApJ, 545, 63

  \bibitem[1992]{emmering92} Emmering, R.T., Blandford, R.D., \& Shlosman, I.
       1992, ApJ, 385, 460

  \bibitem[2001]{ferrarese01} Ferrarese, L., Pogge, R.W., Peterson, B.M. et al.
       2001, ApJ, 555, L79

%  \bibitem[2000]{kaspi00} Kaspi, S., Smith, P.S., Netzer, H. et al.
%       2000, ApJ, 533, 631

  \bibitem[1994]{koenigl94} K\"{o}nigl, A., \& Kartje, J.E.
       1994, ApJ, 434, 446

  \bibitem[2001]{kollatschny01} Kollatschny, W., Bischoff, K., Robinson, E.L.,
      Welsh, W.F., \& Hill, G.J.  2001, A\&A, 379, 125 (Paper\,I)


  \bibitem[2002]{kollatschny02} Kollatschny, W., \& Bischoff, K. 2002,
       A\&A, 386, L19 (Paper\,II)

  \bibitem[1996]{kollatschny96} Kollatschny, W., \& Dietrich, M. 1996,
       A\&A, 314, 43

  \bibitem[1997]{kollatschny97} Kollatschny, W.,\& Dietrich, M. 1997,
       A\&A, 323, 5

  \bibitem[1991]{koratkar91} Koratkar, A., Gaskell, M.  1991, ApJ 370, L61

  \bibitem[2001]{krolik01} Krolik, J.\ H.\ 2001, ApJ, 551, 72

  \bibitem[1997]{murray97} Murray, N., \& Chiang, J.  1997, ApJ, 474, 91


  \bibitem[1994]{obrien94} O'Brien, P.T., Goad, M.R., \& Gondhalekar, P.M.,
         1994, MNRAS, 268, 845

  \bibitem[1992]{perez92} Perez, E., Robinson, A., \& de la Fuente, L.
         1992, MNRAS, 256, 103


  \bibitem[2000]{proga00} Proga, D., Stone, J.M., \& Kallman, T.R.
       2000, ApJ, 543, 686

  \bibitem[1996]{ulrich96} Ulrich, M.-H., \& Horne, K. 1996, MNRAS, 283, 748

  \bibitem[1991]{welsh91} Welsh, W.F., \& Horne, K.  1991, ApJ, 379, 586

  \bibitem[2001]{welsh01} Welsh, W.F. 2001, in: Peterson et al. (eds.): 
       Probing the Physics of AGN, ASP Conf.Ser. 224, p.123

\end{thebibliography}
\end{document}